\documentclass[reprint,superscriptaddress,amsmath,amssymb,pra,aps,showpacs]{revtex4-1}

\usepackage{cases}
\usepackage{amsmath,amsthm}
\usepackage {xcolor}
\usepackage{comment}

\usepackage{mathrsfs}
\usepackage{comment}
\usepackage{amssymb}
\usepackage{graphicx}
\usepackage{subfigure}

\usepackage[colorlinks,
linkcolor=blue,
anchorcolor=blue,
citecolor=blue]{hyperref}

\renewcommand{\thesubsection}{\Alph{subsection}}

\newtheoremstyle{mydefstyle}
{}
{}
{\normalfont}
{2em}
{\bfseries}
{.}
{ }
{}

\theoremstyle{mydefstyle}
\newtheorem{definition}{Definition}
\newtheorem{theorem}{Theorem}
\newtheorem{lemma}{Lemma}

\newtheorem{remark}{Remark}
\newtheorem{proposition}{Proposition}

\usepackage{braket}

\usepackage{graphicx}
\usepackage{dcolumn}
\usepackage{bm}

\begin{document}
	
	\title{Magic resource theory and central limit theorem for quantum channels via Ruzsa divergence }

	\author{Linmao Wang}
	\affiliation{Department of Mathematics, Nanchang University,
		Nanchang 330031, P R China}	
		
			\author{Linlin Ye}
		\affiliation{Department of Mathematics, Nanchang University,
			Nanchang 330031, P R China}	
			
				\author{Zhaoqi Wu}
			\email{wuzhaoqi\_conquer@163.com}
			\affiliation{Department of Mathematics, Nanchang University,
				Nanchang 330031, P R China}

	\author{Chunhe Xiong}
	\affiliation{%
		\resizebox{0.88\textwidth}{!}{%
			School of Mathematics and Statistics, HNP-LAMA,
			Central South University, Changsha 410083, China%
		}%
	} 
	
	\author{Shao-Ming Fei}
		\email{feishm@cnu.edu.cn}
	\affiliation{School of Mathematical Sciences,
		Capital Normal University, Beijing 100048, P R China}
		
	\begin{abstract}
		The celebrated Gottesman-Knill theorem indicates that the magic
		resources are crucial to achieve quantum computational advantage.
		Thus, the characterization of magic resources for quantum states and
		quantum channels has emerged as a key research topic. In this work,
		focusing on multi-qudit systems, we establish a framework of
		magic resource theory in which the free states and free operations
		are defined as the sets of all minimal stabilizer-projection states
		(MSPSs) and completely MSPS-preserving operations (CMPOs),
		respectively. Based on this framework, we propose two magic
		quantifiers for quantum channels in terms of quantum Ruzsa
		divergence via the Choi-Jamio\l kowski isomorphism, and compare them
		with magic measures induced by the Choi divergence. Moreover, we propose two explicit representations of the mean channel and establish
		the quantum central limit theorem (Q-CLT) for quantum channels via
		Ruzsa divergence, which implies that the $N$-fold convolution of a
		zero-mean channel converges to a CMPO and yields a tighter upper bound on the convergence rate compared to previous ones based on other magic quantifiers. The results provide new
		insights into the magic resource theory and quantum central limit
		theorem for quantum channels.
	\end{abstract}
	\maketitle

	\section{Introduction}\label{Sec1}
	The Gottesman-Knill theorem \cite{Gottesman1,Gottesman2,Gottesman3} stands as one of the seminal results in quantum computation. It asserts that any quantum circuit composed solely of Clifford gates can be efficiently simulated on a classical computer in polynomial time \cite{Aaronson,Anders}. Therefore, nonstabilizerness plays an essential role in demonstrating quantum advantage. This motivated the first proposal of the magic resource theory by Veitch $et$ $al.$ \cite{Veith1,Veith2}. Subsequently, numerous magic quantifiers for quantum states and quantum channels have been introduced \cite{Veith1,Heinrich,Liu,Bravyi,Heimendahl,Howard,Leone1,Leone2,Feng,Dai,Bu,Seddon1,Bu1,Wang,Wang2,Saxena,Seddon,Luo}. In the majority of these works, stabilizer states and stabilizer channels (or completely stabilizer-preserving operations) were adopted as free states and free operations, respectively.
	
	The minimal stabilizer-projection state (MSPS), originally defined by Bu
	$et$ $al.$ \cite{Bu2,Bu3}, is the normalized Hermitian operator
	obtained by averaging over the corresponding stabilizer subgroup of
	the generalized Pauli group. Furthermore, it can be viewed as a generalization of the pure stabilizer state, that is, the stabilizer subgroup of an MSPS is not required to be maximal. While
	stabilizer states are conventionally viewed as the discrete-variable
	analogue of continuous-variable Gaussian states from the perspective
	of classical simulability, recent works reveal that when viewed
	through the lens of the quantum central limit theorem (Q-CLT), it is the
	MSPSs that genuinely capture the mathematical elegance of Gaussian
	states \cite{Bu1,Bu2}. This perspective and the elegant properties of
	Gaussian states in continuous-variable
	systems \cite{Bartlett,Mari,Lloyd,Bartlett1,Albarelli,Takagi,Lund,Zhong}
	render the set of all MSPSs a natural candidate for free states in
	magic resource theory within discrete-variable systems. Motivated by
	these facts, we adopt that the set of all MSPSs be free states
	in this work, and further propose the definitions of free operations
	and free superoperations, thereby facilitating the establishment of a
	comprehensive magic resource theory framework.
	
	Ruzsa divergence originates from additive combinatorics and has been
	extended to the entropic framework, which plays a key role in
	entropic inequalities for sum/difference random variables, the
	entropy power inequality and inverse sumset theorems \cite{Tao,Kontoyiannis,Ruzsa,Madiman1}. Recently, Bu $et$ $al$.  \cite{Bu1}
	generalized classical Ruzsa divergence to its quantum
	counterpart based on quantum convolution and von Neumann entropy,
	introduced the quantum Ruzsa divergence of magic and the quantum
	doubling constant, and proposed an entropic Q-CLT where the rate of
	convergence is bounded by the magic gap. Thus, a natural
	idea to quantify magic resource for quantum channels arises via the
	Choi-Jamio\l kowski isomorphism \cite{Choi,Ja}, and the approach has
	been applied in various resource theories including
	nonstabilizerness \cite{Luo,Wang2,Xu,Chen}. On the other hand, Bu $et$ $al$. \cite{Bu2} derived  a Q-CLT for quantum states via the Schatten $L^{2}$ norm and relative entropy, and provided a Q-CLT for quantum channels via the diamond norm. 
	
	In this work, we derive a Q-CLT for quantum channels based on Ruzsa divergence.  In particular, we establish explicit characteristic function and Kraus representations of the mean channel, with the latter constructed from the stabilizer group of the corresponding Choi state. These results provide a concrete characterization of the limiting channel under iterative convolution. We demonstrate that CMPOs are not merely a subset of CSPOs, but serve as the ultimate converging targets under the iterative convolution of quantum channels, playing a role strictly analogous to that of Gaussian channels in continuous-variable systems. This provides a compelling physical motivation for our framework: the Ruzsa divergence explicitly quantifies how the nonstabilizerness (magic) of a channel exponentially decays, ultimately thermalizing into a completely MSPS-preserving operation.

	This work is organized as follows. In Section \ref{Sec2}, we
	introduce frequently used notation and briefly
	review the stabilizer formalism. In Section \ref{Sec3}, we develop a
	novel framework of magic resource theory and further show the
	inclusion relations of the free sets among distinct existing
	frameworks. In Section \ref{Sec4}, we propose the magic quantifiers
	induced by quantum Ruzsa divergence via the Choi-Jamio\l kowski
	isomorphism, and prove that one of them is a magic monotone for
	quantum channels. In Section \ref{Sec5}, we compare our magic
	quantifiers with those induced by the Choi divergence. In Section
	\ref{Sec6}, we derive the characteristic function and the Kraus representations of the mean channel, and then establish a Ruzsa divergence-based Q-CLT for quantum channels. Finally, we conclude in
	Section \ref{Sec7}. 
	
	\section{Preliminaries}\label{Sec2}
	
We focus on multi-qudit systems with the Hilbert space $\mathcal{H}^{\otimes n}$, where $\mathcal{H} \cong \mathbb{C}^{d}$ is the $d$-dimensional Hilbert space for a prime $d$. $\mathbb{Z}_{d}$ is the ring of integers modulo $d$, $\left\lbrace |j\rangle\right\rbrace_{j \in \mathbb{Z}_{d}} $ and $\left\lbrace |\vec{j} \rangle=|j_{1} \rangle \otimes \cdots \otimes |j_{n} \rangle \right\rbrace_{\vec{j} \in \mathbb{Z}_{d}^{n}} $ are the computational bases on $\mathcal{H}$ and $\mathcal{H}^{\otimes n}$, respectively. We conventionally utilize $\mathcal{H}_{A}$ and $\mathcal{H}_{B}$ to denote the Hilbert spaces $\mathcal{H}^{\otimes n_{A}}$ and $\mathcal{H}^{\otimes n_{B}}$ respectively, where $n_{A},n_{B} \in \mathbb{Z}^{+}$ are positive integers.  $\mathcal{L}\left(\mathcal{H}\right)$ is the set of all linear operators on $\mathcal{H}$, and $\mathcal{D}\left(\mathcal{H}\right)$ is the set of all density operators on $\mathcal{H}$. 

A quantum operation $\Phi$ is a completely positive map with the Kraus representation $\Phi(\rho) = \sum_{i} K_{i}\rho K_{i}^{\dagger}$, in which $\{K_{i}\}$ is a set of Kraus operators from $\mathcal{H}_{\mathcal{A}}$ into $\mathcal{H}_{\mathcal{B}}$ satisfying $\sum_{i} K_{i}^{\dagger} K_{i} \leqslant \mathbb{I}$, where $\mathbb{I}$ denotes the identity operator. $\Phi$ is called a quantum channel if it is trace-preserving, i.e., $\sum_{i} K_{i}^{\dagger} K_{i} = \mathbb{I}$. Let $\mathcal{O}_{A \rightarrow B}$ be the set of all quantum operations from $\mathcal{L}\left(\mathcal{H}_{A}\right)$ to $\mathcal{L}\left(\mathcal{H}_{B}\right)$ and $\mathfrak{S}_{AB \rightarrow A^{\prime}B^{\prime}}$ be the set of all quantum superoperations from $\mathcal{O}_{A \rightarrow B}$ to $\mathcal{O}_{A^{\prime} \rightarrow B^{\prime}}$. $\mathcal{C}_{A \rightarrow B}$ denotes all quantum channels from $\mathcal{L}\left(\mathcal{H}_{A}\right)$ to $\mathcal{L}\left(\mathcal{H}_{B}\right)$. Moreover, when considering the composition of two quantum channels, we always assume that the output space of the first channel matches the input space of the second. We denote $\psi=|\psi\rangle\langle\psi|$ the density matrix of a pure state $|\psi \rangle$ for brevity. Furthermore, $\Phi, \Theta,\mathcal{E}$ are quantum operations, $\text{id}_{A}$ is the identity channel on $\mathcal{H}_{A}$, $\Phi \otimes \Theta$ and $\Phi \circ \Theta$ denote the tensor product and the composition of $\Phi$ and $\Theta$, respectively. Finally, $\Xi_{\mathcal{S}},\left\lbrace  \Xi_{\mathcal{S},i}\right\rbrace_{i \in \mathbb{Z}^{+}} $ denote superoperations.

	The Pauli $X$ and $Z$ operators on the Hilbert space $\mathcal{H}$ are defined as \cite{Schwinger}
	\[
	X=\sum_{j=0}^{d-1} |j+1 \rangle \langle j |,~~~ Z=\sum_{j=0}^{d-1} \chi\left(j\right) |j\rangle \langle j|,
	\]
where $\chi\left(j\right)=\omega^{j}$, with $\omega=\text{e}^{\frac{2\pi \text{i}}{d}}$ being the primitive $d$-th root of unity in which $\text{i}$ is the imaginary unit. 

The local Weyl operators (generalized Pauli operators) are defined as \cite{Appleby}
	\[
	w\left(p,q\right)=
	\begin{cases}
		\chi \left(-2^{-1}pq\right) Z^{p} X^{q}, \,\ d>2,\\
		\text{i}^{-pq} Z^{p} X^{q},\,\ d=2,
	\end{cases}
	\]
	where $p,q \in \mathbb{Z}_{d}$ and $2^{-1}$ denotes the inverse of $2$ in $\mathbb{Z}_{d}$, i.e., $\frac{d+1}{2}$.
	For any $d>2$, the Weyl operators satisfy
	\[
	w\left(p,q\right) w\left(p^{\prime},q^{\prime}\right)=	\chi \left(2^{-1} \left\langle\left(p,q\right),\left(p^{\prime},q^{\prime}\right) \right\rangle_{s} \right) w\left(p+p^{\prime}, q+q^{\prime}\right),
	\]where $\left\langle \cdot, \cdot \right\rangle_{s} $ is the symplectic inner product given by $\left\langle\left(p,q\right),\left(p^{\prime},q^{\prime}\right) \right\rangle_{s}=pq^{\prime}-qp^{\prime}.$
	Consequently, it follows that $\left[w\left(p,q\right), w\left(p^{\prime},q^{\prime}\right)\right]=0$ if and only if $\left\langle\left(p,q\right),\left(p^{\prime},q^{\prime}\right) \right\rangle_{s}=0$, which provides an equivalent condition for the commutativity of the Weyl operators.
	
	For a composite system on the Hilbert space $\mathcal{H}^{\otimes n}$, the Weyl operators can be written as $w\left(\vec{p},\vec{q}\right)=w\left(p_{1},q_{1}\right) \otimes \cdots \otimes w\left(p_{n},q_{n}\right)$,
	where $\vec{p}=\left(p_{1}, \cdots , p_{n}\right) \in \mathbb{Z}_{d}^{n}$, $\vec{q}=\left(q_{1}, \cdots , q_{n}\right) \in \mathbb{Z}_{d}^{n}$.  Denote $\vec{x}=\left(\vec{p},\vec{q}\right) \in V^{n} \equiv \mathbb{Z}_d^n \times \mathbb{Z}_d^n$. The Weyl operators $w\left(\vec{x}\right)$ form an orthonormal basis for all linear operators on $\mathcal{H}^{\otimes n}$ with respect to the Hilbert-Schmidt inner product $\left\langle A, B\right\rangle =\frac{1}{d^{n}} \text{Tr} (A^{\dagger} B)$. Hence, any quantum state $\rho \in \mathcal{D}\left(\mathcal{H}^{\otimes n}\right)$ can be written as a linear combination of Weyl operators, $	\rho=\frac{1}{d^{n}} \sum_{\vec{x} \in V^{n}} \Xi_{\rho}\left(\vec{x}\right) w\left(\vec{x}\right)$, where $\Xi_{\rho}\left(\vec{x}\right)=\text{Tr} \left[\rho w\left(-\vec{x}\right)\right]$ is the characteristic function of $\rho$. Furthermore, the mean state of $\rho$ is defined as \cite{Bu3}
	\begin{equation}\label{eq1}
		\mathcal{M}\left(\rho\right)=\frac{1}{d^{n}} \sum_{\vec{x} \in V^{n}: |\Xi_{\rho}\left(\vec{x}\right)|=1} \Xi_{\rho}\left(\vec{x}\right) w\left(\vec{x}\right).
	\end{equation}
The transformation to the characteristic function is the quantum Fourier transform, which has a wide range of applications in the discrete Hudson theorem, quantum Boolean functions and quantum circuit complexity \cite{Gross,Jaffe}. Moreover, the magic gap of $\rho$ is defined as \cite{Bu3}
	\[
	MG(\rho)=1-\max_{\vec{x}: 0 < |\Xi_{\rho}\left(\vec{x}\right)| <1}  |\Xi_{\rho}\left(\vec{x}\right)|,
	\]
	which denotes the gap between the largest absolute value, which is 1, and the second-largest absolute value in the support of the characteristic function.
	
	The generalized Pauli group consisting of Weyl operators with phase factors is defined as
	\[
	\mathcal{P}=\left\lbrace \chi\left(j\right) w\left(\vec{x}\right)\middle| j\in \mathbb{Z}_{d}, \vec{x} \in V^{n} \right\rbrace .
	\]
	The Clifford group $\mathcal{CL}$ is the normalizer of $\mathcal{P}$ \cite{Veith1}, i.e.,
	\[
	\mathcal{CL}=\left\lbrace U:  Uw\left(\vec{x}\right)U^{\dagger} \in \mathcal{P}, \,\ \forall \vec{x} \in V^{n} \right\rbrace.
	\]
		A pure stabilizer state $\psi \in \mathcal{D}\left( \mathcal{H}^{\otimes n}\right) $ is defined as a common eigenstate corresponding to eigenvalue +1 of a maximal abelian subgroup $\mathcal{S}_{\psi }$ of size $d^{n}$. If the generators of the abelian subgroup of $\mathcal{P}$ are $g_{1}, \cdots, g_{n}$, it holds that \cite{Bu1}
	\[
	|\psi\rangle \langle \psi|=\frac{1}{d^{n}} \sum_{g \in \mathcal{S}_{\psi }} g,
	\]
	where $g=g_{1}^{k_{1}} \cdots g_{n}^{k_{n}}$, $k_{i} \in \mathbb{Z}_{d}$. We denote the set of all pure stabilizer states by $\mathcal{PS}$, and $\mathcal{STAB}$ the set of all stabilizer states, which is the convex hull of $\mathcal{PS}$.
	
	The stabilizer group of a quantum state $\rho \in \mathcal{D}\left(\mathcal{H}^{\otimes n}\right)$ is defined as
	\[
	\mathcal{S}_{\rho}=\left\lbrace g : g \in \mathcal{P}, g\rho=\rho \right\rbrace.
	\]
	As an example, $\mathcal{S}_{|\vec{0}\rangle \langle \vec{0}|}=\left\langle Z_{1}, \cdots, Z_{n} \right\rangle$, where $Z_{i}$ is the Pauli $Z$ operator on the $i$-th qudit.
	Consider a general abelian subgroup $\mathcal{S}_\rho = \langle g_1, \cdots, g_r \rangle \subseteq \mathcal{P}$ of size $d^r$, where $r \in [n]= \{0, 1, \cdots, n\}$. A quantum state $\rho \in \mathcal{D}(\mathcal{H}^{\otimes n})$ is called a minimal stabilizer-projection state (MSPS) associated with $\mathcal{S}_\rho$ if it takes the form \cite{Bu2},
	\[
	\rho=\frac{1}{d^{n}} \sum_{g \in \mathcal{S}_{\rho}} g,
	\]
	where $g=g_{1}^{k_{1}} \cdots g_{r}^{k_{r}}$, $k_{i} \in \mathbb{Z}_{d}$. We denote by $\mathcal{MSPS}$ the set of all minimal stabilizer-projection states. Equivalently \cite{Bu3}, for a given abelian subgroup $\mathcal{S} \subseteq \mathcal{P}$, one can define the associated stabilizer projections as all projections within $C^*(\mathcal{S})$, the $C^*$-algebra generated by $\mathcal{S}$. The MSPSs are then precisely the normalized minimal stabilizer projections.
	
	Note that no extra dimensional subscript is used in the definition of sets $\mathcal{P},\mathcal{CL},\mathcal{MSPS},$ etc. In this work, we consistently assume automatic dimension matching and thus adopt a unified notation. For instance, if $w\left(\vec{x}\right) \in \mathcal{P}$ and $\rho  \in \mathcal{MSPS}$ are the operators on $\mathcal{H}^{\otimes n}$, the sets implicitly denote $\mathcal{P} \cap \mathcal{L}\left(\mathcal{H}^{\otimes n}\right)$ and $\mathcal{MSPS} \cap \mathcal{D}\left(\mathcal{H}^{\otimes n}\right)$. Moreover, for the sets of quantum channels in the following sections, we also adhere to the same convention.

	\section{Framework of magic resource theory}\label{Sec3}
	
In the common existing frameworks of magic resource theory, the free states are mainly treated as the stabilizer states \cite{Veith1,Howard,Leone1,Leone2}, or  the set of subnormalized states with nonpositive mana $\mathcal{W}$ defined as \cite{Wang}
	\[
	\mathcal{W} := \left\{ \sigma : \sum_{\mathbf{u}} |W_{\sigma}(\mathbf{u})| \leqslant 1, \sigma \geqslant 0 \right\},
	\] 
where $W_{\sigma}(\mathbf{u})=\frac{1}{d}\text{Tr}\left[A_{\mathbf{u}}\sigma\right]$ is the discrete Wigner function. Frequently used free operations are the stabilizer operations ($\mathcal{SO}$) which consist of the following types of quantum operations: Clifford operations, tensoring in stabilizer states,
	partial trace, measurements in the computational basis, and post-processing conditioned on these measurement results \cite{Veith1,Leone2,Bu1} or completely stabilizer-preserving operations ($\mathcal{CSPO}$) \cite{Saxena} defined as the set of quantum operations $\Phi$ satisfying
$	\left( \text{id}_{R} \otimes \Phi \right) \left(\rho\right) \in \mathcal{STAB}$ for any $\rho \in \mathcal{STAB}$ and ancillary system $R$. In addition, free operations are occasionally regarded as  completely positive Wigner-preserving operations ($\mathcal{CPWPO}$) \cite{Wang2} or stabilizer-preserving operations ($\mathcal{SPO}$), which are those that
	transform stabilizer states to stabilizer states, and which form the largest set of physical operations that can be considered free for the resource theory of nonstabilizerness. Note that any stabilizer channel $\mathcal{E} \in \mathcal{C}_{A \rightarrow A}$ can be represented by the Stinespring representation,
	\begin{equation}\label{eq2}
		\mathcal{E}\left(\rho\right)=\text{Tr}_{B} \left[V \left(\rho \otimes \sigma\right) V^{\dagger}\right],
	\end{equation}
	where $\sigma \in \mathcal{STAB}$ and $V \in \mathcal{L}\left( \mathcal{H}_{A} \otimes \mathcal{H}_{B}\right) $ is a Clifford unitary.
	
The free superoperations conventionally correspond to the relationship between free states and free operations. For instance, if the free operations are $\mathcal{CSPO}$, one may consider free superoperations as (completely) $\mathcal{CSPO}$-preserving superoperations.
	
We introduce a distinct framework of magic resource theory  by adopting $\mathcal{MSPS}$ as free states. We show that $\mathcal{MSPS}$ is a proper subset of $\mathcal{STAB}$ and strictly contains $\mathcal{PS}$ in the following proposition, 	see proof in Appendix \ref{Appendix A}.

	\begin{proposition}\label{pro1}
		For quantum states on $\mathcal{H}^{\otimes n}$, it holds that
		\begin{equation}\label{eq3}
			\mathcal{PS} \subsetneq \mathcal{MSPS} \subsetneq \mathcal{STAB} \subsetneq \mathcal{W}.
		\end{equation}
	\end{proposition}
	
It follows directly from Proposition \ref{pro1} that $\text{Conv}\left(\mathcal{MSPS}\right)=\mathcal{STAB}$, where $\text{Conv}\left(\mathcal{MSPS}\right)=\left\lbrace \sum_{i} p_{i} \rho_{i}:  \rho_{i} \in \mathcal{MSPS}, p_{i} \geq 0, \sum_{i}p_{i}=1 \right\rbrace$ is the convex hull of $\mathcal{MSPS}$. Moreover, our set of free states is strictly smaller than those adopted in prior literature.

For any $\rho \in \mathcal{MSPS}$ and ancillary system $R$ with Hilbert space $\mathcal{H}_{R}$, we define $\mathcal{MPO}$ and $\mathcal{CMPO}$ to be the sets of quantum operations $\Phi \in \mathcal{O}_{A \rightarrow B}$ which satisfy $\Phi \left(\rho\right) \in \mathcal{MSPS}$ and $\left( \text{id}_{R} \otimes \Phi \right) \left(\rho\right) \in \mathcal{MSPS}$, respectively. In the following, we adopt $\mathcal{CMPO}$ as the set of free operations. 

As shown in \cite{Seddon}, for any quantum channel $\Phi \in \mathcal{C}_{A^{\prime} \rightarrow B}$, it holds that
$J_{\Phi} \in \mathcal{STAB} \Leftrightarrow \Phi \in \mathcal{CSPO}$, where $J_{\Phi}$ is the Choi state of $\Phi$ defined as \cite{Choi,Ja}
	\[
	J_{\Phi}=\left(\text{id}_{A} \otimes \Phi\right) \left(|\Omega^{AA^{\prime}} \rangle \langle \Omega^{AA^{\prime}}| \right),
	\]
	which can be equivalently expressed as
	\[
	J_{\Phi}=\frac{1}{d^{n_{A}}} \sum_{\vec{j}, \vec{j'}  \in \mathbb{Z}_{d}^{n_{A}} } |\vec{j} \rangle \langle \vec{j}^{\prime}| \otimes \Phi\left(|\vec{j}\rangle \langle \vec{j}^{\prime}|\right).
	\]
	Here $|\Omega^{AA^{\prime}} \rangle=\sqrt{\frac{1}{d^{n_{A}}}} \sum_{\vec{j} \in \mathbb{Z}_{d}^{n_{A}}} |\vec{j} \vec{j} \rangle$ is the maximally entangled state on $\mathcal{H}_{A} \otimes \mathcal{H}_{A^{\prime}}$, where $\mathcal{H}_{A} \cong \mathcal{H}_{A^{\prime}}$. Moreover, for any input state $\rho \in \mathcal{D}\left(\mathcal{H}^{\otimes n_{A}}\right)$, it holds that
	\begin{equation}\label{eq4}
		\Phi \left(\rho\right)=d^{n_{A}} \text{Tr}_{A} \left[J_{\Phi} \left(\rho^{T} \otimes \mathbb{I}_{B}\right)\right].
	\end{equation}
Motivated by this equivalence, we establish the following theorem, see proof in Appendix \ref{Appendix B}.
	
	\begin{theorem}\label{thm1}
		For any quantum channel $\Phi \in \mathcal{C}_{A \rightarrow B}$, it holds that
		\begin{equation}\label{eq5}
			J_{\Phi} \in \mathcal{MSPS} \Leftrightarrow \Phi \in \mathcal{CMPO}.
		\end{equation}
	\end{theorem}
	
	Theorem \ref{thm1} provides a state-level equivalent condition for whether a channel belongs to $\mathcal{CMPO}$. Based on Theorem \ref{thm1}, the relation among various free operations is given by the following proposition, see proof  in Appendix \ref{Appendix C}.

	\begin{proposition}\label{pro2}
		For quantum operations from $\mathcal{H}_{A}$ to $\mathcal{H}_{B}$, it holds that
		\begin{equation}\label{eq6}
			\mathcal{CMPO} \subsetneq \mathcal{CSPO}  \subsetneq \mathcal{SPO}.
		\end{equation}
	\end{proposition}
	
\begin{figure}[ht]
		\centering
		\begin{minipage}{0.7\linewidth}
			\centering
			\includegraphics[width=\textwidth]{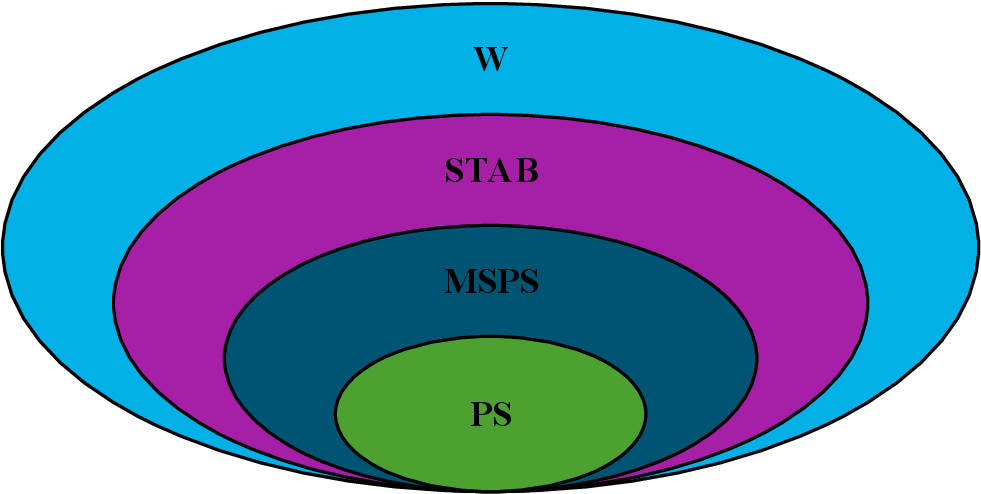}\\
			(a)
		\end{minipage}
		\hfill
		\begin{minipage}{0.7\linewidth}
			\centering
			\includegraphics[width=\textwidth]{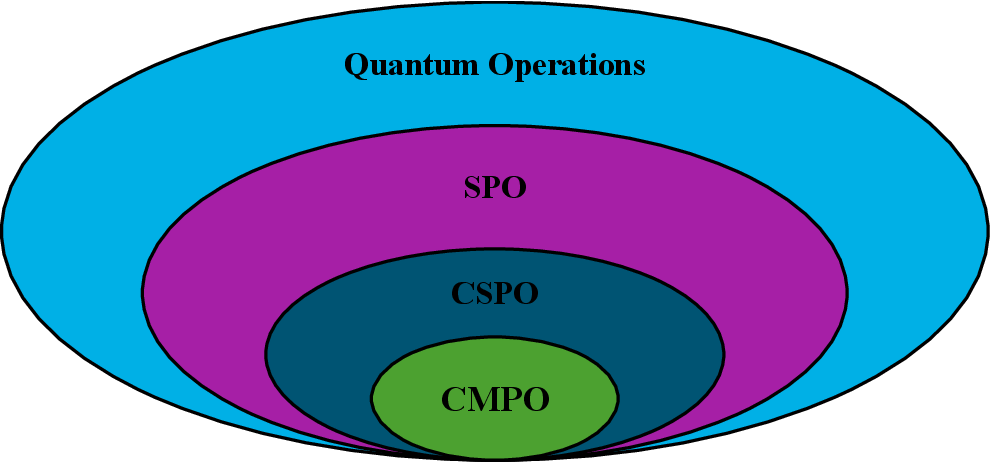}\\
			(b)
		\end{minipage}
		
		\vspace{0.3cm}
		\caption{Illustration of the relations among different free states and free operations in magic resource theory. (a) Hierarchy of various free states; (b) Hierarchy of various free operations.}
		\label{fig1}
	\end{figure}

The above proposition indicates that the set of our free operations is strictly smaller than existing ones, mirroring the hierarchy established for free states. Figure \ref{fig1} depicts two Venn diagrams illustrating the relation between free states and free operations for distinct frameworks.
	
Although $\mathcal{CMPO}$ forms a strictly smaller subset of free operations compared to existing frameworks, this stringent definition is mathematically and physically indispensable. As we shall demonstrate in Section \ref{Sec6}, CMPOs naturally emerge as the asymptotic limits in the quantum central limit theorem for channels, endowing them with a unique operational significance akin to Gaussian channels in continuous-variable systems.
		
Given the established frameworks of free states and free operations, we proceed to present the definition of free superoperations motivated by the Stinespring representation of stabilizer channels and the frameworks in coherence and imaginarity resource theories for quantum channels \cite{Theurer,Wu}.
	
	\begin{definition}\label{def2}
		For any free operation $\mathcal{E} $ and Clifford channels  $\mathcal{E}_{U}$ and $\mathcal{E}_{V}$, the elementary free superoperations are given by
		\[
		\Xi_{\mathcal{S},1}\left(\Phi\right)= \mathcal{E} \otimes \Phi, \,\ \Xi_{\mathcal{S},2}\left(\Phi\right)= \Phi \otimes \mathcal{E} ,
		\]
		\[
		\Xi_{\mathcal{S},3}\left(\Phi\right)= \mathcal{E}_{U} \circ \Phi, \,\ \Xi_{\mathcal{S},4}\left(\Phi\right)= \Phi \circ \mathcal{E}_{V} ,
		\]
		and
		\[
		\Xi_{\mathcal{S},5}\left(\Theta\right)= \mathcal{T}_{R} \circ \Theta,
		\]
		where we assume $\Phi \in \mathcal{O}_{A \rightarrow B} ,\Theta \in \mathcal{O}_{A \rightarrow BR}$ and $\mathcal{T}_{R}$ is the partial trace over the ancillary system $R$.
		
		A superoperation $\Xi_{\mathcal{FS}}$ is free if and only if it can be written as a sequence of elementary free superoperations,
		\[
		\Xi_{\mathcal{FS}}=\Xi_{\mathcal{S},i_{n}} \circ \cdots \circ \Xi_{\mathcal{S},i_{2}} \circ
		\Xi_{\mathcal{S},i_{1}},
		\]
		where $i_j\in \{1,2,3,4,5\}$, $j\in \{1,2,\cdots,n\}$ and $n\in
		\mathbb{Z}^{+}$.
	\end{definition}
We have the following conclusion, see proof in Appendix \ref{Appendix D}.
	
	\begin{theorem}\label{thm2}
For any free channel $\Phi$,  $\Xi_{\mathcal{FS}}\left(\Phi\right)$ is a free channel.
	\end{theorem}
	
From Theorem \ref{thm2}, our free superoperations indeed preserve the set of free operations. Moreover, the magic measure of quantum operations can be regarded as a high-level concept of the magic measure of quantum states, in the sense that a magic measure of quantum operations is a functional $M$ which maps a quantum operation to $\left[0,+\infty\right)$ satisfying the following properties:
	
	(i) Faithfulness: $M\left(\Phi\right)=0$ iff $\Phi$ is a free operation for $\Phi \in \mathcal{O}_{A \rightarrow B}$;
	
	(ii) Monotonicity under free superoperations: $M\left(\Xi_{\mathcal{FS}} \left[\Phi\right]\right)\le M\left(\Phi\right),$ where $\Phi \in \mathcal{O}_{A \rightarrow B}$ and $\Xi_{\mathcal{FS}} \in \mathfrak{S}_{AB \rightarrow A^{\prime}B^{\prime}}$;
	
	(iii) Convexity: $M\left(\sum_{i} p_{i} \Phi_{i}\right) \leq \sum_{i} p_{i} M\left(\Phi_{i}\right)$,
	where $\left\lbrace \Phi_{i}\right\rbrace_{i} \subseteq \mathcal{O}_{A \rightarrow B} $ and $\left\lbrace p_{i} \right\rbrace_{i} $ is a classical probability distribution.
	
	We say that $M$ is a magic monotone of quantum operations if (i) and (ii) hold.
	
\section{ Magic quantifiers for quantum channels}\label{Sec4}
	
We next introduce the definition of quantum Ruzsa divergence for quantum channels by applying the Choi-Jamio\l kowski isomorphism and  present the magic quantifiers for quantum channels based on the quantum Ruzsa divergence. Given two $n$-qudit quantum systems associated with Hilbert spaces $\mathcal{H}_{A}$ and $\mathcal{H}_{B}$ (with $n_A = n_B = n$), consider the $2 \times 2$ parameter matrix,
	\[
	G=
	\begin{bmatrix}
		g_{00} & g_{01}\\
		g_{10} & g_{11}
	\end{bmatrix},
	\]
	where at most one of the entries is zero and the determinant $\det(G) \not\equiv 0 \mod d$. The quantum convolution of two quantum states $\rho \in \mathcal{D}\left(\mathcal{H}_{A}\right)$ and $\sigma \in \mathcal{D}\left(\mathcal{H}_{B}\right)$ is defined as \cite{Bu2}
	\[
	\rho \boxtimes \sigma=\text{Tr}_{B} \left[U \left(\rho \otimes \sigma\right) U^{\dagger}\right],
	\]
	where $U$ is the key unitary given by
	\[
	U=\sum_{\vec{i},\vec{j}} |N g_{11} \vec{i}-Ng_{10} \vec{j} \rangle \langle \vec{i}| \otimes |-N g_{01} \vec{i}+Ng_{00} \vec{j} \rangle \langle \vec{j}|,
	\]
	in which $|\vec{i} \rangle=|i_{1} \rangle \otimes \cdots \otimes |i_{n} \rangle \in \mathcal{H}^{\otimes n}$ and $N=\left[ \text{det}\left(G\right)\right] ^{-1} \in \mathbb{Z}_{d}$. The quantum convolution satisfies the following convolution-multiplication duality \cite{Bu3},
	\begin{equation}\label{eq7}
		\Xi_{\rho \boxtimes \sigma}\left(\vec{p},\vec{q}\right)=\Xi_{\rho}\left(Ng_{11}\vec{p},g_{00}\vec{q}\right) \Xi_{\sigma}\left(-Ng_{10}\vec{p},g_{01}\vec{q}\right)
	\end{equation}
	for any $\vec{p},\vec{q} \in \mathbb{Z}_{d}^{n}$.
	
	Unless otherwise specified, we adopt the convention that the convolution $\boxtimes$ is $\boxtimes_{s,t}$, the discrete beam splitter with the parameter matrix  \cite{Bu1},
	\[
	G_{s,t}=
	\begin{bmatrix}
		s & t\\
		t & -s
	\end{bmatrix},
	\]
	where $s^{2}+t^{2} \equiv 1 \,\ \text{mod} \,\ d$. Moreover, it is noteworthy that this convolution exists nontrivially only for local system dimensions $d \geq 7$. For convenience, we continue to utilize the notations $U$ and $\boxtimes$ whenever we consider the key unitary and the discrete beam splitter. And we denote $\boxtimes^{N+1}\rho = (\boxtimes^N \rho) \boxtimes \rho$,  $\boxtimes^0 \rho = \rho$ \cite{Bu1}.
	
Given any quantum convolution $\boxtimes$, the quantum Ruzsa divergence of a state $\rho \in \mathcal{D}\left( \mathcal{H}^{\otimes n}\right) $ with respect to the state $\sigma \in \mathcal{D}\left( \mathcal{H}^{\otimes n}\right) $ is defined as \cite{Bu1}
	\[
	D_{Rz}\left(\rho \middle\| \sigma\right)=S \left(\rho \boxtimes \sigma\right)-S \left(\rho\right),
	\]
	where $S\left(\rho\right)=-\text{Tr}(\rho \text{log} \rho)$ is the von Neumann entropy of $\rho$.
The quantum Ruzsa divergence of magic $M_{Rz}$ and the quantum-doubling constant $\delta_{q}$ of a state $\rho \in \mathcal{D}\left( \mathcal{H}^{\otimes n}\right)$ are defined as \cite{Bu1}
	\[
	M_{Rz} \left(\rho\right)=\min_{\sigma \in  \mathcal{STAB}} D_{Rz}\left(\rho \middle\| \sigma\right),
	\]
	and
	\[
	\delta_{q}\left(\rho\right)=\exp \left\lbrace S\left(\rho \boxtimes \rho\right)-S\left(\rho\right)\right\rbrace,
	\]
	respectively. It is shown that $M_{Rz}$ and $\delta_{q}$ are both magic monotones within the framework of $\mathcal{MSPS}$ as the set of free states and $\mathcal{SO}$ as the set of free operations.
	
	Motivated by these state-level concepts, we now introduce the quantum Ruzsa divergence for quantum channels via the Choi-Jamio\l kowski isomorphism. Given any quantum convolution $\boxtimes$, we define the quantum Ruzsa divergence between quantum channels $\Phi, \Theta \in \mathcal{C}_{A \rightarrow B}$ as
		\begin{equation}\label{eq8}
			\mathcal{D}_{Rz} \left(\Phi \middle\| \Theta\right):=S \left(J_{\Phi} \boxtimes J_{\Theta}\right)-S \left(J_{\Phi}\right).
		\end{equation}
$\mathcal{D}_{Rz}$ has the following properties, see proof in Appendix \ref{Appendix E}.

	\begin{lemma}\label{lemma1}
		For quantum channels $ \left\lbrace \Phi_{i} \right\rbrace_{i}, \left\lbrace \Theta_{i} \right\rbrace_{i} \subseteq \mathcal{O}_{A \rightarrow B} $,
		
		(i) \textbf{Positivity:} $\mathcal{D}_{Rz} \left(\Phi \middle\| \Theta\right) \geq 0$ and the equality holds iff $J_{\Phi} \in C^{*}\left(\mathcal{S}_{J_{\Theta}}\right)$. Moreover, $\mathcal{D}_{Rz} \left(\Phi \middle\| \Phi \right) =0$ iff $\Phi \in \mathcal{CMPO}$.
		
		(ii) \textbf{Additivity under tensor product:}
		\[
		\mathcal{D}_{Rz}\left(\Phi_{1} \otimes \Phi_{2} \middle\| \Theta_{1} \otimes \Theta_{2} \right)=\mathcal{D}_{Rz} \left(\Phi_{1} \middle\| \Theta_{1} \right)+\mathcal{D}_{Rz} \left(\Phi_{2} \middle\| \Theta_{2}\right).
		\]
		
		(iii) \textbf{Invariance under Clifford channels:}
		\[
		\mathcal{D}_{Rz} \left(\mathcal{E}_{U} \circ \Phi  \middle\| \mathcal{E}_{U} \circ \Theta \right)=\mathcal{D}_{Rz} \left(\Phi \circ \mathcal{E}_{V}  \middle\| \Theta \circ \mathcal{E}_{V} \right)=\mathcal{D}_{Rz} \left(\Phi \middle\| \Theta\right),
		\]
		where $\mathcal{E}_{U},\mathcal{E}_{V}$ are Clifford channels.
		
		(iv) \textbf{Convexity (concavity) on the first (second) channel:}
		\[
		\mathcal{D}_{Rz} \left(\sum_{i} p_{i} \Phi_{i} \middle\| \Theta \right) \leq \sum_{i} p_{i} \mathcal{D}_{Rz} \left(\Phi_{i} \middle\| \Theta \right),
		\]
		\[
		\mathcal{D}_{Rz} \left( \Phi \middle\| \sum_{i} q_{i} \Theta_{i} \right) \geq \sum_{i} q_{i} \mathcal{D}_{Rz} \left(\Phi \middle\| \Theta_{i} \right),
		\]
		where $\left\lbrace p_{i} \right\rbrace_{i}, \left\lbrace q_{i} \right\rbrace_{i}$ are classical probability distributions.
	\end{lemma}
	
Due to the concavity of $D_{Rz}\left(\rho \middle\| \sigma\right)$ with respect to $\sigma$ and the definition of $\mathcal{STAB}$, it follows that $ \min_{\sigma \in  \mathcal{STAB}}  D_{Rz}\left(\rho \middle\| \sigma\right)$ achieves the minimal value at some pure stabilizer state, which implies that
	\[
	\min_{\sigma \in  \mathcal{STAB}} D_{Rz}\left(\rho \middle\| \sigma\right)=\min_{\sigma \in  \mathcal{PS}} D_{Rz}\left(\rho \middle\| \sigma\right).
	\]
	Furthermore, we have
	\begin{align}\label{eq9}
		\min_{\sigma \in  \mathcal{STAB}} D_{Rz}\left(\rho || \sigma\right)
		=&\min_{\sigma \in  \mathcal{MSPS}} D_{Rz}\left(\rho \middle\| \sigma\right)\notag\\
		=& \min_{\sigma \in  \mathcal{PS}} D_{Rz}\left(\rho \middle\| \sigma\right)
	\end{align}
	owing to  Proposition \ref{pro1}. Accordingly, for any quantum channel $\Phi \in \mathcal{C}_{A \rightarrow B}$, we define the quantum Ruzsa divergence of magic for $\Phi$ as
		\begin{equation}\label{eq10}
			\mathcal{M}_{Rz} \left(\Phi\right):=\min_{\Theta \in \mathcal{CMPO}} \mathcal{D}_{Rz} \left(\Phi \middle\| \Theta\right).
		\end{equation}
	
It is worth noting that the state-level equivalence in Eq. (\ref{eq9}) does not generally extend to quantum channels. Indeed, the Choi state of a quantum channel must satisfy the additional marginal constraint \(\operatorname{Tr}_B J_\Theta=\mathbb{I}_A/d^{n_A}\). Consequently, the minimization over $\mathcal{CMPO}$ in Eq. (\ref{eq10}) cannot, in general, be replaced by that over $\mathcal{CSPO}$. $\mathcal{M}_{Rz}$ has the following desirable properties, see proof in Appendix \ref{Appendix F}.
	
	\begin{theorem}\label{thm3}
	$\mathcal{M}_{Rz}$ is monotonic under free superoperations.
	\end{theorem}
	
	\begin{remark} \label{remark1}
		Note that the faithfulness of $\mathcal{M}_{Rz}$ holds only in one direction. That is, if $\Phi \in \mathcal{CMPO}$, we have
		\[
		0 \leq \mathcal{M}_{Rz}\left(\Phi\right) = \min_{\Theta \in \mathcal{CMPO}} \mathcal{D}_{Rz} \left(\Phi \middle\| \Theta\right) \leq \mathcal{D}_{Rz} \left(\Phi \middle\| \Phi\right)=0,
		\]
		where the second inequality follows by choosing the specific free channel $\Theta \in \mathcal{CMPO}$, thus we have $\mathcal{M}_{Rz}\left(\Phi\right) =0$ from Lemma \ref{lemma1} (i).
		
Suppose that $\mathcal{M}_{Rz}\left(\Phi\right) =0$. Then there exists a free channel $\Theta^{*}$ such that $\mathcal{D}_{Rz} \left(\Phi \middle\| \Theta^{*}\right)=0 $, indicating that $J_{\Phi} \in C^{*}\left( \mathcal{S}_{J_{\Theta^{*}}}\right) $. Hence, $J_{\Phi} \in \mathcal{STAB}$ implies $\Phi \in \mathcal{CSPO}$, and $\mathcal{M}_{Rz}$ is not a genuine magic monotone for quantum channels. Nevertheless, if $\Phi$ is a unitary channel, we have $\mathcal{M}_{Rz}\left(\Phi\right)=0$ iff $\Phi \in \mathcal{CMPO}$, which follows from the fact that $J_{\Phi}$ is a pure state iff $\Phi $ is a unitary channel.
	\end{remark}
	
Given a quantum channel $\Phi \in \mathcal{C}_{A \rightarrow B}$, we correspondingly define the logarithmic quantum-doubling constant by,
		\begin{equation}\label{eq11}
			\mathcal{LM}_{\delta}\left(\Phi\right):= S\left(J_{\Phi} \boxtimes J_{\Phi}\right)-S\left(J_{\Phi}\right).
		\end{equation}
Utilizing Lemma \ref{lemma1} and Corollary 3 in \cite{Bu1}, by imitating the proof of Theorem \ref{thm3}, we have the following theorem.
	
	\begin{theorem}\label{thm4}
		$\mathcal{LM}_{\delta}$ is a magic monotone for quantum channels.
	\end{theorem}
	
	\begin{remark}\label{remark2}
		Theorem \ref{thm4} shows that $\mathcal{LM}_{\delta}$ is a genuine and computable magic monotone. Moreover, it reduces to the quantum-doubling constant up to logarithm, i.e.,
		\[
		\mathcal{LM}_{\delta}\left(\Phi_{\text{re}}\right)=\log\left[ \delta_{q}\left(\sigma\right)\right] ,
		\]
		where $\Phi_{\text{re}}$ is the replacement channel which acts as $\Phi_{\text{re}}\left(\rho\right)=\text{Tr}\left[\rho\right] \sigma$. In fact, it holds that
		\begin{align*}
			&\mathcal{LM}_{\delta}\left(\Phi_{\text{re}}\right)\\
			=&  S\left(J_{\Phi_{\text{re}}} \boxtimes J_{\Phi_{\text{re}}}\right)-S\left(J_{\Phi_{\text{re}}}\right)\\
			=& \ S\left[ \left( \frac{\mathbb{I}_{A}}{d^{n_{A}}} \otimes \sigma\right)  \boxtimes  \left( \frac{\mathbb{I}_{A}}{d^{n_{A}}} \otimes \sigma\right) \right]-S \left( \frac{\mathbb{I}_{A}}{d^{n_{A}}}\otimes \sigma\right)  \\
			=&  S\left[\left(\frac{\mathbb{I}_{A}}{d^{n_{A}}} \boxtimes \frac{\mathbb{I}_{A}}{d^{n_{A}}}\right) \otimes \left(\sigma \boxtimes \sigma\right)\right] -S \left(\frac{\mathbb{I}_{A}}{d^{n_{A}}} \otimes \sigma\right) \\
			=& S\left(\frac{\mathbb{I}_{A}}{d^{n_{A}}}\right)+ S\left(\sigma \boxtimes \sigma\right)-S\left(\frac{\mathbb{I}_{A}}{d^{n_{A}}}\right)-S\left(\sigma\right) \\
			=& \log \left[ \delta_{q}\left(\sigma\right)\right] ,
		\end{align*}
		where the second equality comes from the fact that $J_{\Phi_{\text{re}}}=\frac{\mathbb{I}_{A}}{d^{n_{A}}} \otimes \sigma$, the third follows from the identity $\left(  \rho_{1} \otimes \rho_{2}\right) \boxtimes \left(\sigma_{1} \otimes \sigma_{2}\right)=\left(\rho_{1} \boxtimes \sigma_{1} \right) \otimes \left(\rho_{2} \boxtimes \sigma_{2} \right)$, and the penultimate one follows from the equations $\frac{\mathbb{I}_{A}}{d^{n_{A}}} \boxtimes \frac{\mathbb{I}_{A}}{d^{n_{A}}}=\frac{\mathbb{I}_{A}}{d^{n_{A}}}$ and $S\left(\rho \otimes \sigma\right)=S\left(\rho\right)+S\left(\sigma\right)$.
	\end{remark}
	
We now explicitly demonstrate the precise computability of our magic monotone by evaluating some fundamental physical noise models in single-qudit systems with odd prime dimension $d \geq 7$.
	
	\textbf{Example 1.} Consider the depolarizing channel $\mathcal{D}_p \in \mathcal{O}_{A^{\prime} \rightarrow B}$ \cite{Nielsen},
	\[
	\mathcal{D}_p(\rho) = (1-p)\rho + p\frac{\mathbb{I}_d}{d}.
	\]
	By direct calculations, we obtain
	\[
	J_{\mathcal{D}_p}= (1-p)\left(|\Omega^{AA^{\prime}} \rangle \langle \Omega^{AA^{\prime}}| \right)+ p \frac{\mathbb{I}_{d^2}}{d^2},
	\]
	and
	\[
	J_{\mathcal{D}_p} \boxtimes J_{\mathcal{D}_p} = J_{\mathcal{D}_{2p-p^2}}.
	\]	
	Thus, it holds that
	\[
	\mathcal{LM}_\delta(\mathcal{D}_p)=\frac{d^2-1}{d^2} \int_{p}^{2p-p^2} \log \left( \frac{(d^2-1)(1-x)+1}{x} \right) \text{d}x.
	\]
	
\begin{figure}
		{\begin{minipage}[figure2]{0.88\linewidth}
				\includegraphics[width=1.03\textwidth]{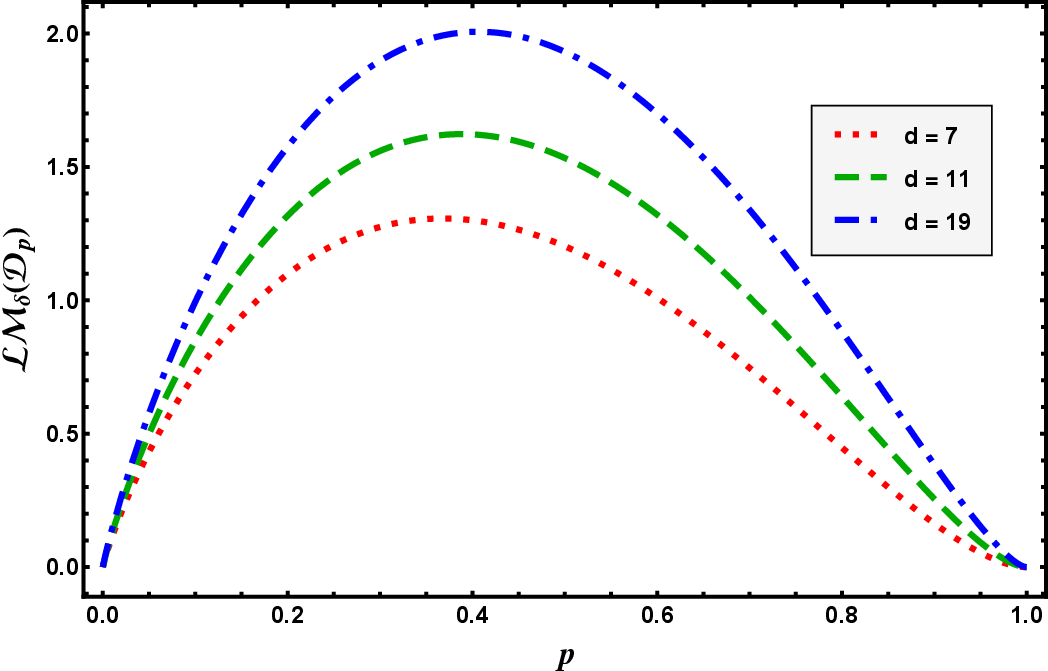}
		\end{minipage}}
		\caption{{Behavior of the channel magic monotone $\mathcal{LM}_\delta(\mathcal{D}_p)$ under varying depolarizing channels as functions of $p$ with different dimensions: $d=7$ (red dotted line); $d=11$ (green dashed line); $d=19$ (blue dot dashed line). }}
		\label{figure2}
	\end{figure}

	We depict $\mathcal{LM}_\delta(\mathcal{D}_p)$ in Figure \ref{figure2}, which clearly visualizes the nonlinear variation of magic resources across continuous noise regimes for various dimensions. Notice that for any $d$ and $p$, there exist only two CMPOs, i.e., the identity channel at $p=0$, and the completely depolarizing channel at $p=1$. Moreover, as \(d\) increases, both the maximum value of \(\mathcal{LM}_\delta(\mathcal{D}_p)\) and the corresponding maximizing \(p\) increase, where the former grows asymptotically as $\frac{\log d}{2}$ and the latter approaches $\frac{1}{2}$.
		
	\textbf{Example 2.} Consider the Werner-Holevo channel \cite{Werner},
	\[
	\Phi_{\text{WH}}\left(\rho\right)=\frac{1}{d-1}\left(\mathbb{I}_d -\rho^T\right),
	\]
	which gives a counterexample to an additivity conjecture for output purity of quantum channels, where $\rho^T$ denotes the transpose of $\rho$ in the computational basis. 
	
	A straightforward calculation yields
	\[
	J_{\Phi_{\text{WH}}} = \frac{1}{d(d-1)} (\mathbb{I}_{d^2} - S),
	\]
	where $S= \sum_{i,j} \vert{}i\rangle\langle j\vert{} \otimes \vert{}j\rangle\langle i\vert{}$ is the swap operator, and
	\[
	J_{\Phi_{\text{WH}}} \boxtimes  J_{\Phi_{\text{WH}}}= \frac{d-2}{d(d-1)^2} \mathbb{I}_{d^2} + \frac{1}{d(d-1)^2} S.
	\]	
	Thus, it holds that
	\[
	\mathcal{LM}_\delta\left(\Phi_{\text{WH}}\right)=1 + \frac{d-3}{2(d-1)} \log_2 \left( \frac{d-1}{d-3} \right).
	\]
	
\begin{figure}
		{\begin{minipage}[figure2]{0.88\linewidth}
				\includegraphics[width=1.03\textwidth]{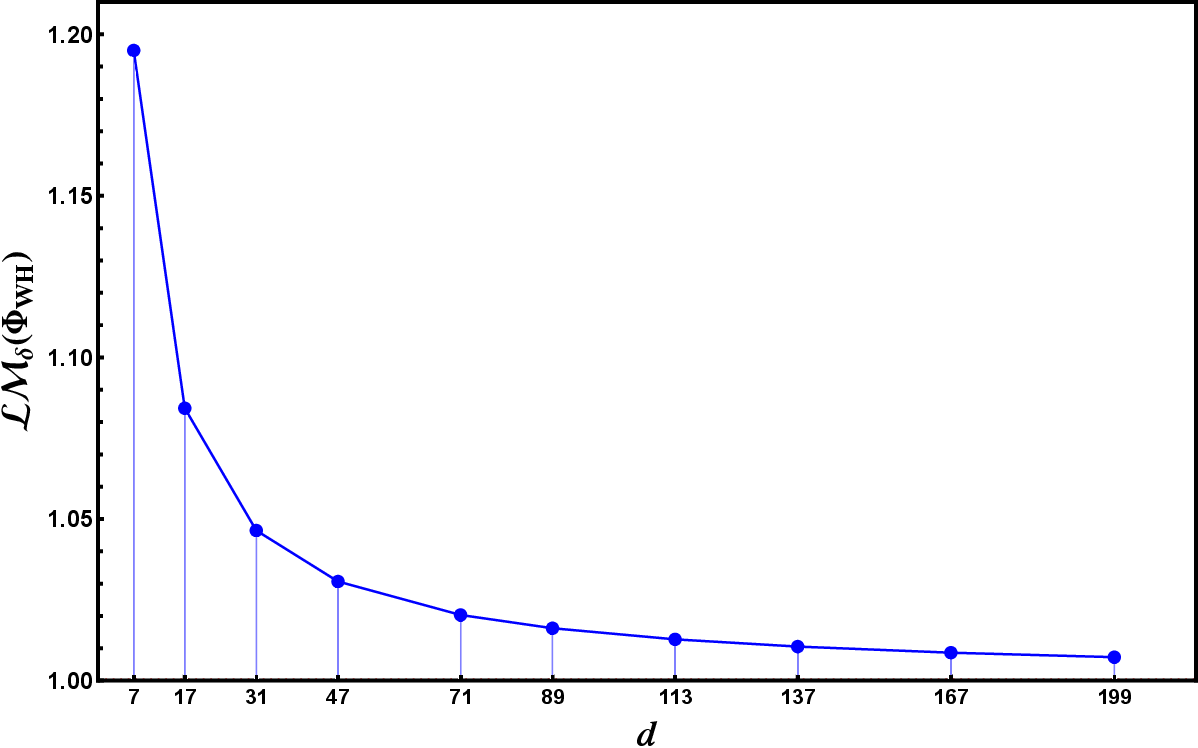}
		\end{minipage}}
		\caption{{Behavior of the channel magic monotone $\mathcal{LM}_\delta(\Phi_{\text{WH}})$ under varying odd prime dimension $d$. }}
		\label{figure3}
	\end{figure}

We depict $\mathcal{LM}_\delta(\Phi_{\text{WH}})$ in Figure \ref{figure3}, which shows that magic monotone decreases strictly monotonically to 1 as the odd prime dimension $d \geq 7$ increases. This implies that the Werner-Holevo channel is strictly resourceful, that is, it never reduces to a free operation, regardless of the system dimension.
	
	\textbf{Example 3.} For $s \equiv \pm t$, consider the unitary channel $U_T$ induced by quantum $T$-gate \cite{Gottesman},
	\[
	T_d=\sum_{j=0}^{d-1} \omega^{j^3} |j\rangle\langle j|,
	\]
	which is vital in quantum circuit for achieving quantum advantage.
	By direct calculations, it holds that
	\[
	J_{U_T}=|T_d \rangle \langle T_d|,
	\]
	and 
	\[
	J_{U_T} \boxtimes J_{U_T}= \frac{1}{d^2} \sum_{y=0}^{d-1} \sum_{x, x'} \omega^{P(x,y) - P(x',y)} \vert{}x, x\rangle\langle x', x'\vert{},
	\]
	where $	\vert{}T_d \rangle = \frac{1}{\sqrt{d}} \sum_{j=0}^{d-1} \omega^{j^3} \vert{}j j\rangle$ and $P(x,y) = (sx+ty)^3 + (tx-sy)^3$.
	
\begin{figure}
		{\begin{minipage}[figure2]{0.88\linewidth}
				\includegraphics[width=1.03\textwidth]{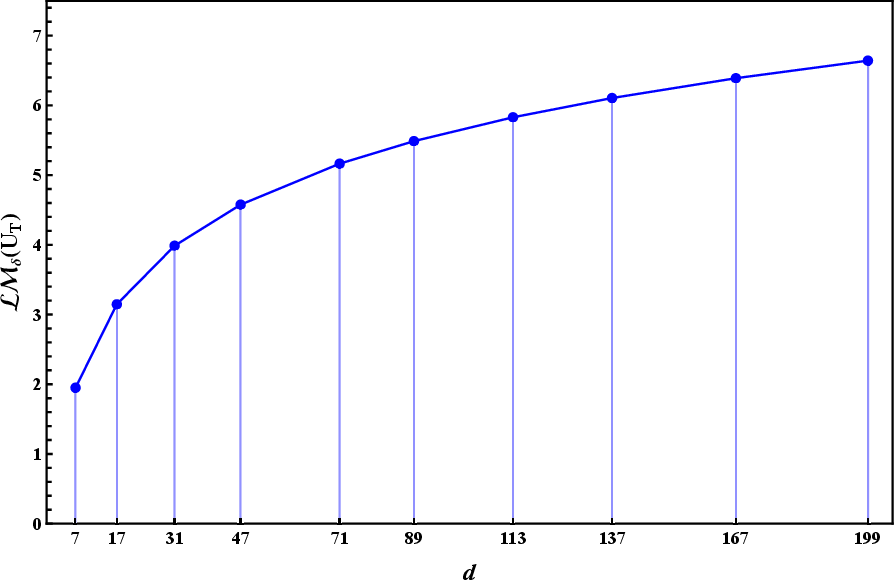}
		\end{minipage}}
		\caption{{Behavior of the channel magic monotone $\mathcal{LM}_\delta(U_T)$ under varying odd prime dimension $d$.}}
		\label{figure4}
	\end{figure}

	Evaluating the spectrum via Gauss sums reveals that the eigenvalues of $J_{U_T} \boxtimes J_{U_T}$ are $\frac{2}{d}$, $\frac{1}{d}$ and 0, with multiplicities $\frac{d-1}{2}$, $1$ and $d^2-\frac{d+1}{2}$, respectively. Hence, noting that $S(J_{U_T}) = 0$, we have
	\[
	\mathcal{LM}_{\delta}\left(U_T\right)=S\left(J_{U_T} \boxtimes J_{U_T}\right)=\log d-\frac{d-1}{d}.
	\]
	We depict $\mathcal{LM}_\delta(U_T)$ in Figure \ref{figure4}. In stark contrast to the Werner-Holevo channel (Figure \ref{figure3}), the magic monotone of $U_T$ exhibits a strictly monotonic increase with respect to the odd prime dimension $d \geq 7$. Furthermore, a direct comparison between Example 2 and Example 3 reveals a profound resource gap, that is, the unitary channel induced by quantum $T$-gate possesses strictly and significantly greater magic resources than the Werner-Holevo channel $\Phi_{\text{WH}}$, highlighting the crucial role of $T$-gate in enabling universal quantum computation.
		
	\section{Bounds for magic monotones}\label{Sec5}
	
We analyze in this section the relationship between our magic quantifiers and those induced by the Choi divergence \cite{Gilad}.
To begin with, inspired by the definitions of the relative entropy of magic \cite{Veith1} and the modified relative entropy of magic \cite{Bu4}, we define the magic measures induced by the Choi divergence as follows.
For any quantum channel $\Phi \in \mathcal{C}_{A \rightarrow B}$, we define the  Choi divergence of magic and its modified version as
		\begin{equation}\label{eq12}
			\mathcal{M}_{R}\left(\Phi\right):=\min_{\Theta \in \mathcal{CSPO}} \mathcal{D}_{R}\left(\Phi \middle\| \Theta\right),
		\end{equation}
		and
		\begin{equation}\label{eq13}
			\widetilde{\mathcal{M}}_{R}\left(\Phi\right):=\min_{\Theta \in \mathcal{CMPO}} \mathcal{D}_{R}\left(\Phi \middle\| \Theta\right),
		\end{equation}
		respectively, where $\mathcal{D}_{R}\left(\Phi \middle\| \Theta\right)=D\left(J_{\Phi} \middle\| J_{\Theta}\right)$ is the Choi divergence, with $D\left(\cdot \middle\| \cdot\right)$ being the quantum relative entropy.
	
	Since $\mathcal{CMPO} \subsetneq \mathcal{CSPO}$, it immediately follows that $\mathcal{M}_R(\Phi) \leq \widetilde{\mathcal{M}}_R(\Phi)$ for any $\Phi \in \mathcal{O}_{A \rightarrow B}$. Moreover, one can verify that $	\mathcal{M}_{R}$ and $	\widetilde{\mathcal{M}}_{R}$ are magic measures for quantum channels, similar to the relative entropy of magic. The following proposition establishes the exact relationship between $\mathcal{M}_R$ and our Ruzsa divergence-based magic quantifier $\mathcal{M}_{Rz}$, see proof in Appendix \ref{Appendix G}.
	
	\begin{proposition}\label{pro3}
		For any quantum channel $\Phi \in \mathcal{C}_{A \rightarrow B}$, it holds that
		\begin{equation}\label{eq14}
			\mathcal{M}_{R}\left(\Phi\right) \leq \mathcal{M}_{Rz}\left(\Phi\right).
		\end{equation}
	\end{proposition}
	
	\begin{proposition}\label{pro4}
		For any quantum channel $\Phi \in \mathcal{C}_{A \rightarrow B}$, we have
		\begin{align}\label{eq15}
			\mathcal{LM}_{\delta}\left(\Phi\right)  \leq \widetilde{\mathcal{M}}_{R}\left(\Phi\right).
		\end{align}
		Moreover, for $s \equiv \pm t$, the equality holds iff $\Phi \in \mathcal{CMPO}$.
	\end{proposition}
	
	\textbf{Proof.} For any $\Phi \in \mathcal{O}_{A \rightarrow B}$, it holds that
	\begin{align*}
		& \mathcal{LM}_{\delta}\left(\Phi\right)\\
		=& S\left(J_{\Phi} \boxtimes J_{\Phi}\right)-S\left(J_{\Phi}\right)\\
		\leq & S\left(\mathcal{M}\left(J_{\Phi} \boxtimes J_{\Phi}\right)\right)-S\left(J_{\Phi}\right)\\
		=& S\left(\mathcal{M}\left(J_{\Phi} \right)\right)-S\left(J_{\Phi}\right)\\
		=& 	\widetilde{\mathcal{M}}_{R}\left(\Phi\right),
	\end{align*}
	where the inequality comes from the fact that for any state $\rho$, 
	$S\left(\rho\right) \leq S\left(\mathcal{M}\left(\rho\right)\right)$, the penultimate equality holds because $S\left(\mathcal{M}\left(J_{\Phi} \boxtimes J_{\Phi}\right)\right)=S\left(\mathcal{M}\left(J_{\Phi} \right)\right)$ and the last equality holds because for any state $\sigma $ \cite{Bu2},
$\min_{\sigma \in \mathcal{MSPS}} D\left(\rho \middle\| \sigma\right)=S\left(\mathcal{M}\left(\rho\right)\right)-S\left(\rho\right)$.
	Hence the equality holds iff $J_{\Phi} \boxtimes J_{\Phi} \in \mathcal{MSPS}$ iff $J_{\Phi} \in \mathcal{MSPS} $ when $s \equiv \pm t$. Then the conclusion follows from Theorem \ref{thm1}. $\hfill\qedsymbol$
	
	\section{Q-CLT: A Ruzsa divergence perspective}\label{Sec6}
	
In continuous-variable systems, the central limit theorem describes a ubiquitous thermalization process where iterative physical processes wash out non-Gaussianity. In this section, we establish the discrete-variable channel analogue of this phenomenon by utilizing the quantum Ruzsa divergence.
	
	Given two $n$-qudit channels $\Phi_1, \Phi_2 \in \mathcal{C}_{A \rightarrow B}$,
	the convolution $\Phi_1 \boxtimes \Phi_2$ is the quantum channel with the Choi state \cite{Bu3},
	$J_{\Phi_1 \boxtimes \Phi_2} := J_{\Phi_1} \boxtimes J_{\Phi_2}$,
	where the state $J_{\Phi_1} \boxtimes J_{\Phi_2}$ is the convolution of the Choi states $J_{\Phi_1}$ and $J_{\Phi_2}$.
	For a given quantum channel $\Phi \in \mathcal{C}_{A \rightarrow B}$, the mean channel $\widetilde{\mathcal{M}}(\Phi)$ is the quantum channel with the Choi state \cite{Bu3},
$J_{\widetilde{\mathcal{M}}(\Phi)}= \mathcal{M}(J_\Phi)$,
	where the state $\mathcal{M}(J_\Phi)$ is the mean state of the Choi state $J_\Phi$. $\Phi$ is called a zero-mean channel if $J_{\Phi}$ is a zero-mean state, which implies that
$\Xi_{\mathcal{M}(J_\Phi)}\left(\vec{x}\right) \in \left\lbrace 0,1 \right\rbrace$	for any $\vec{x} \in V^{n_A +n_B}$.
	
	While the concept of a mean state is well-established, the explicit analytical form of its channel counterpart has remained unexplored in the existing literature. To fill this structural gap, we establish the first concrete formulations of the mean channel. Specifically, for any given quantum channel $\Phi \in \mathcal{O}_{A \to B}$, its corresponding mean channel $\widetilde{\mathcal{M}}(\Phi)$ admits two explicit, equivalent characterizations.
	
	\begin{figure}
		{\begin{minipage}[figure2]{0.92\linewidth}
				\includegraphics[width=0.92\textwidth]{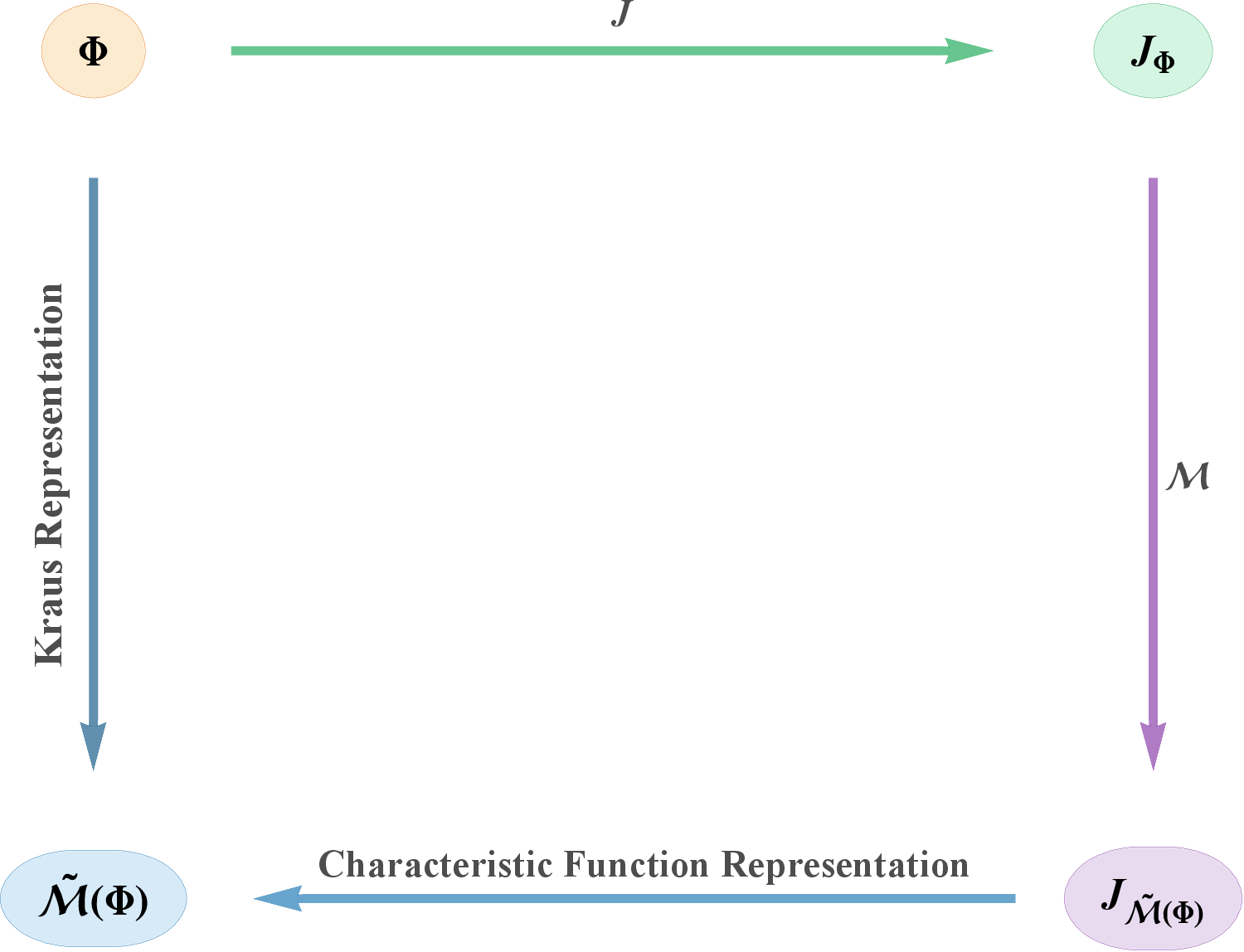}
		\end{minipage}}
		\caption{{ Evolution path of two mean channel representations.}}
		\label{figure5}
	\end{figure}
		
		First,  by applying Eqs. (\ref{eq1}) and (\ref{eq4}), for any quantum channel $\Phi \in \mathcal{C}_{A \rightarrow B}$, the mean channel of $\Phi$ acts as
		\begin{align}\label{eq17}
			&\widetilde{\mathcal{M}}(\Phi)(\cdot)\notag\\
			=& \frac{1}{d^{n_B}} \sum_{ \vert{}\Xi_{J_\Phi}(\vec{x}_A, \vec{x}_B)\vert{} = 1} \Xi_{J_\Phi}(\vec{x}_A, \vec{x}_B) \text{Tr} \big[(\cdot)^T w(\vec{x}_A) \big] w(\vec{x}_B),
		\end{align}
		which is completely determined by the stabilizer group of $J_\Phi$, mirroring the definition of the mean state.
		
		Second, let us establish a direct Kraus representation of the mean channel. To this end, let \(\{(w(\vec x_i),w(\vec y_i),\zeta_i)\}_{i=1}^r\) specify an independent and complete set of the preserved Weyl relations $\Phi^\dagger\left( w\left( \vec y_i\right) \right)=\zeta_i w\left( \vec x_i\right) $, where $\Phi^\dagger$ denotes the adjoint map of $\Phi$, given by $\Phi^\dagger (\cdot) =\sum_{j=1}K_j^{\dagger} (\cdot) K_j$ and $\zeta_i=\omega^{k_i}, k_{i} \in \mathbb{Z}_{d}$. Here, the independence means that the associated operators $g_i=\zeta_i^{-1}\left(\overline{w(\vec x_i)}\otimes w(\vec y_i)\right)$ satisfy $\prod_{i=1}^r g_i^{t_i}=\mathbb{I}$ only when $t_i \equiv 0 \mod d$, and the completeness means that every preserved Weyl relation corresponds to an operator of the form \(g=\prod_{i=1}^{r}g_i^{t_i}\) for some \(t_i\in\mathbb Z_d\). Define
		\[
		\mathcal P_\Phi=\prod_{i=1}^{r} \left(\frac1d\sum_{t=0}^{d-1}\mathcal C_i^t\right),~~
		\mathcal C_i(\cdot)	=\zeta_i^{-1} w\left( \vec y_i\right) (\cdot) w\left( \vec x_i\right) ^\dagger.
		\]
We have the following theorem, see proof in Appendix \ref{Appendix H}.
		
		\begin{theorem}\label{thm5}
			 For any given quantum channel $\Phi \in \mathcal{C}_{A \to B}$ with Kraus representation $	\Phi(\cdot)=\sum_{j=1}K_j(\cdot)K_j^\dagger$, its mean channel $\widetilde{\mathcal{M}}(\Phi)$ admits the Kraus representation
\begin{equation}\label{eq18}
		\widetilde M(\Phi)(\cdot)
		=
		\sum_{\vec a\in\mathbb Z_d^{n_A},\vec b\in\mathbb Z_d^{n_B}}
		\widetilde K_{\vec b,\vec a}
		(\cdot)\widetilde K_{\vec b,\vec a}^{\dagger},
\end{equation}
where $	\widetilde K_{\vec{b},\vec{a}}
=\sqrt{d^{r-n_B}}\,
\mathcal P_\Phi\left( |\vec{b}\rangle\langle \vec{a}|\right) $. 
		\end{theorem}
		
		Figure \ref{figure5} provides a visual schematic of these two representations. The Kraus construction provides a complementary representation of the mean channel. For channels with additional algebraic structure, this representation can be further simplified, facilitating the analysis of their noise dynamics. As an example, consider the bit-flip channel in single-qudit system with odd prime dimension $d=7$ \cite{Nielsen},
$\Phi(\rho) = (1 - p)\rho + pX\rho X^\dagger$. For $0<p<1$, its mean channel can be expressed as
\[
\widetilde{ \mathcal{M}}(\Phi)(\rho)
=\sum_{k=0}^{d-1}\widetilde K_k\rho\widetilde K_k^\dagger,
\]
where
\[
	\widetilde K_k=|\widetilde k\rangle\langle\widetilde k|, \quad |\widetilde k\rangle=\frac1{\sqrt d}\sum_{j=0}^{d-1}\omega^{jk}|j\rangle,
\]
in which the states \(\left\lbrace |\widetilde k\rangle\right\rbrace_k \) form the Fourier basis.
		
Now we give the Q-CLT for quantum channels via Ruzsa divergence, see proof in Appendix \ref{Appendix I}.
	
	\begin{theorem}\label{thm6}
		Given an $n$-qudit zero-mean channel $\Phi $, $N$-fold convolution $\boxtimes^{N} \Phi$ converges to a completely MSPS-preserving channel  when $N \rightarrow \infty$. More precisely, it holds that
		\begin{align}\label{eq19}
			&	\mathcal{D}_{Rz}\left(\boxtimes ^{N} \Phi \middle\| \widetilde{\mathcal{M}}\left(\Phi\right)\right) \notag\\
			\leq & \log \left\lbrace 1+ \frac{1}{d^r} \sum_{\vec{x}: w\left(\vec{x}\right) \notin \mathcal{S}_{\Phi}} \vert{}\Xi_{J_\Phi}(\vec{x})\vert{}^{2N+2} \right\rbrace,
		\end{align}
		where $\mathcal{S}_{\Phi}$ is the corresponding stabilizer group of $J_{\Phi}$ with order $d^r$.
	\end{theorem}
		
Note that the Q-CLT for channels induced by quantum Ruzsa divergence yields a strictly tighter upper bound than those induced by the diamond norm or quantum relative entropy derived in \cite{Bu1,Bu3}. We leave the detailed proof in Appendix \ref{Appendix J}.
	
\begin{figure}
		\centering
		\begin{minipage}{0.88\linewidth}
			\centering
			\includegraphics[width=\textwidth]{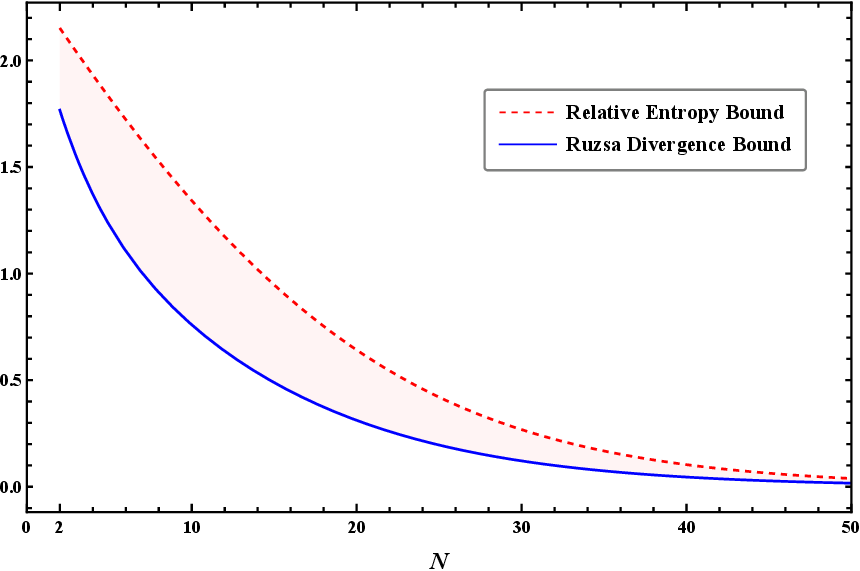}\\
			(a)
		\end{minipage}
		\hfill
		\begin{minipage}{0.88\linewidth}
			\centering
			\includegraphics[width=\textwidth]{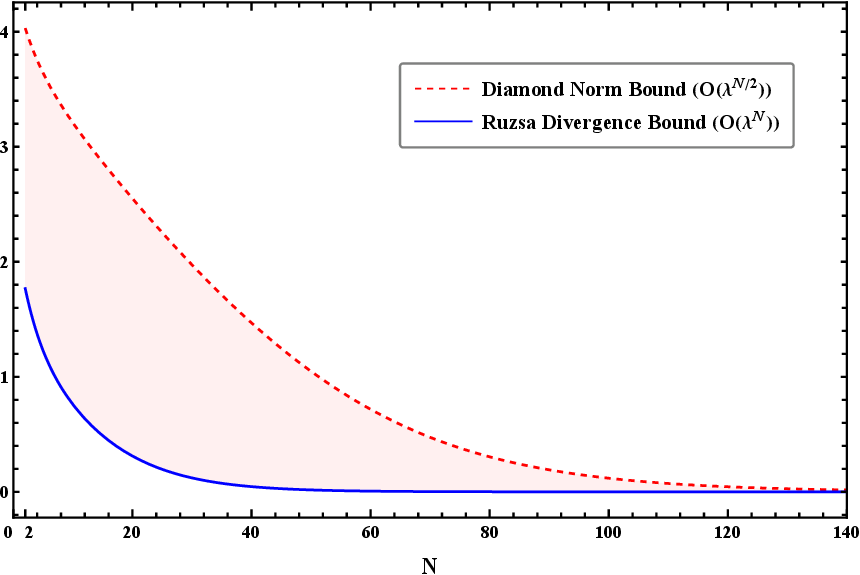}\\
			(b)
		\end{minipage}
		\vspace{0.2cm}
		\caption{Illustration of the asymptotic upper bounds for the quantum central limit theorem of quantum channels under the single-qudit phase-flip channel with $d=7$ and $p=0.15$. (a) Comparison between the Ruzsa divergence bound and the relative entropy bound; (b) Comparison between the Ruzsa divergence bound and the Pinsker inequality-based diamond norm bound, with the latter plotted on a $\log_2(1+x)$ scale.}
		\label{fig6}
	\end{figure}

To explicitly visualize this theoretical advantage, we evaluate and compare the bounds derived in Appendix \ref{Appendix J} in single-qudit system with odd prime dimension $d=7$ under the phase-flip channel \cite{Nielsen},
$\Phi(\rho) = (1-p)\rho + p Z \rho Z^\dagger$. Figure \ref{fig6} depicts the comparative results for $p = 0.15$. As indicated by the red shaded area, the significant gap intuitively demonstrates that our bound is tighter. Furthermore, comparing the slopes of the curves clearly shows that our bound exhibits a faster asymptotic convergence rate.
	
Although quantum Ruzsa divergence is not a metric, Theorem \ref{thm6} demonstrates that it can nevertheless be used to establish a quantum central limit theorem for quantum channels. Furthermore, it is worth noting that the role of CMPO in the Q-CLT is equivalent to the Gaussian distribution of the central limit theorem in probability theory. Therefore, within the framework of quantum channel convolutions, CMPO emerges as an indispensable structural element, offering profound insights for future research.
	
	\section{Summary}\label{Sec7}
	
We have established a new framework of magic resource theory. According to this framework, we have proposed the magic quantifiers for quantum channels induced by quantum Ruzsa divergence of magic via the Choi–Jamiołkowski isomorphism. Our comparative analysis reveals a clear theoretical hierarchy: $\mathcal{LM}_\delta$ provides a rigorous lower bound for $\widetilde{\mathcal{M}}_R$, while $\mathcal{M}_{Rz}$ serves as an upper bound for $\mathcal{M}_R$. Moreover, as a foundational result, we have derived two explicit, equivalent representations of the mean channel for any given quantum channel. Moreover, we have demonstrated that under iterative convolutions, the nonstabilizerness of a zero-mean channel exponentially decays, ultimately thermalizing into a completely MSPS-preserving operation. Notably, our Ruzsa divergence-induced bound scales as $O(\lambda^N)$, making it strictly tighter than the existing approaches based on the diamond norm or quantum relative entropy. We have obtained several fundamental results, shedding new light on the study of magic resource theory and quantum central limit theorem.
	
	\begin{acknowledgments}
The core ideas and overall approach of this work were developed by the authors. During manuscript preparation, the authors used generative AI tools to assist with language editing, the derivation of Theorem \ref{thm5}, as well as checking and reformulation of some derivations. The authors independently verified and revised all AI-assisted mathematical content and take full responsibility for the results and the final manuscript. This work was supported by National Natural Science Foundation of China (Grant Nos. 12561084,12161056) and Natural Science Foundation of Jiangxi Province of China (Grant No. 20232ACB211003).
	\end{acknowledgments}
	
	\subsection*{Data Availability}
	No datasets were generated or analysed during the current study.
	
	\appendix
	\renewcommand{\thesubsection}{\Alph{subsection}}
	
	\begin{widetext}
		\refstepcounter{subsection}
		\subsection*{Appendix \thesubsection. Proof of Proposition \ref{pro1}}\label{Appendix A}
		
		Initially, $\mathcal{STAB} \subsetneq \mathcal{W}$ is the result of \cite{Wang}. By setting $r=0$ and $r=n$, we obtain $\mathcal{PS} \subsetneq \mathcal{MSPS}$. Thus we only need to show that $\mathcal{MSPS} \subsetneq \mathcal{STAB}$.
		
		For any quantum state $\rho \in \mathcal{MSPS}$,  denote the corresponding stabilizer group by
		\[
		\mathcal{S}_{\rho}=\left\langle g_{1},g_{2},\cdots ,g_{r}\right\rangle, \,\
		r<n.
		\]
		Then the minimal stabilizer projection $P_{\rho}$ associated with $\mathcal{S}_{\rho}$ takes the following form
		\[
		P_{\rho}=\frac{1}{|\mathcal{S}_{\rho}|} \sum_{g \in \mathcal{S}_{\rho}}^{} g.
		\]
		One can determine that the rank of $P_{\rho}$ is $d^{n-r}$, as
		\begin{align}
			\text{Tr}\left(\frac{1}{| \mathcal{S}_{\rho}|} \sum_{g \in  \mathcal{S}_{\rho}}^{} g\right)=\frac{1}{| \mathcal{S}_{\rho}|}\left( \sum_{g \neq \mathbb{I}}^{} \text{Tr} g+\text{Tr} \mathbb{I}\right)=\frac{d^{n}}{| \mathcal{S}_{\rho}|}=d^{n-r}.
			\tag{A1} \label{eqA1}
		\end{align}
		Moreover, it holds that
		\begin{align}
			\frac{1}{| \mathcal{S}_{\rho}|} \sum_{g \in  \mathcal{S}_{\rho}}^{} g |\psi_{\vec{m}}\rangle=|\psi_{\vec{m}}\rangle,
			\tag{A2} \label{eqA2}
		\end{align}
		for all $\vec{m} \in \mathbb{Z}_{d}^{n-r}$, where $|\psi_{\vec{m}}\rangle$ is the common eigenstate of $g \in  \mathcal{S}_{\rho}$ with corresponding eigenvalue +1.
		
		Combining Eqs. (\ref{eqA1}) and (\ref{eqA2}), we have
		\begin{align}
			\rho=\frac{P_{\rho}}{\text{Tr} \left[P_{\rho}\right]} =\frac{1}{d^{n-r}} \sum_{\vec{m} \in \mathbb{Z}_{d}^{n-r}}^{} |\psi_{\vec{m}} \rangle \langle \psi_{\vec{m}}|,
			\tag{A3} \label{eqA3}
		\end{align}
		where $|\psi_{\vec{m}} \rangle \in \mathcal{PS}$ for all $\vec{m} \in \mathbb{Z}_{d}^{n-r}$. This implies that $\rho \in \mathcal{STAB}$ and has the form of convex combination of pure stabilizer states with equal weights, and thus we have $\mathcal{MSPS} \subseteq \mathcal{STAB}$. Furthermore, we obtain that $\mathcal{MSPS} \subsetneq \mathcal{STAB}$ since $\mathcal{STAB}$ is a convex set but $\mathcal{MSPS}$ is not.  $\hfill\qedsymbol$
		
		\refstepcounter{subsection}
		\subsection*{Appendix \thesubsection.  Proof of Theorem \ref{thm1}}\label{Appendix B}

		For necessity, we divide the proof into two parts. In the first part, we establish the following assertion:
		\begin{align}
			J_{\Phi} \in \mathcal{MSPS} \Rightarrow \Phi \in \mathcal{MPO}.
			\tag{B1} \label{eqB1}
		\end{align}
		
		Without loss of generality, we omit the phase factors of Weyl operators according to Lemma 2 in \cite{Sun}.
		
		For any $\rho \in \mathcal{MSPS}$, we have $\rho^{T} \in \mathcal{MSPS}$ since $\mathcal{S}_{\rho}^{T}$ is a stabilizer subgroup whenever $\mathcal{S}_{\rho}$ is a stabilizer subgroup. If $J_{\Phi} \in \mathcal{MSPS}$,
		let
		\[
		\rho^{T} =\frac{1}{d^{n_{A}}} \sum_{w\left(\vec{x}\right) \in \mathcal{S}_{\rho^{T}}} w\left(\vec{x}\right),
		\]
		and
		\[
		J_{\Phi} =\frac{1}{d^{n_{A}+n_{B}}} \sum_{w\left(\vec{y}\right) \otimes w\left(\vec{z}\right) \in \mathcal{S}_{J_{\Phi}}} w\left(\vec{y}\right) \otimes w\left(\vec{z}\right),
		\]
		where $\mathcal{S}_{\rho^{T}}$ and $\mathcal{S}_{J_{\Phi}}$ are the stabilizer groups of $\rho^{T}$ and $J_{\Phi}$ respectively.
		
		Applying the relationship between $J_{\Phi}$ and $\Phi$, we obtain that
		\begin{align}
			\Phi\left(\rho\right)
			= & d^{n_{A}}\text{Tr}_{A} \left[ \left(\frac{1}{d^{n_{A}+n_{B}}} \sum_{w\left(\vec{y}\right) \otimes w\left(\vec{z}\right) \in \mathcal{S}_{J_{\Phi}}} w\left(\vec{y}\right) \otimes w\left(\vec{z}\right) \right) \left( \frac{1}{d^{n_{A}}}\sum_{w\left(\vec{x}\right) \in \mathcal{S}_{\rho^{T}}} w\left(\vec{x}\right) \otimes \mathbb{I}_{B} \right) \right]\notag\\
			=&\left( \frac{1}{d^{n_{A}+n_{B}}} \sum_{w\left(\vec{y}\right) \otimes w\left(\vec{z}\right) \in \mathcal{S}_{J_{\Phi}}}  \sum_{w\left(\vec{x}\right) \in \mathcal{S}_{\rho^{T}}} \text{Tr} \left[w\left(\vec{y} \right) w\left(\vec{x}\right) \right]  \right) \otimes w\left(\vec{z}\right)\notag\\
			= & \left( \frac{1}{d^{n_{B}}}  \sum_{w\left(\vec{y}\right) \otimes w\left(\vec{z}\right) \in \mathcal{S}_{J_{\Phi}}}\sum_{w\left(\vec{x}\right) \in \mathcal{S}_{\rho^{T}}} \delta_{\vec{x},-\vec{y}}  \right)  \otimes w\left(\vec{z}\right)\notag\\
			=& \frac{1}{d^{n_{B}}} \sum_{w\left(\vec{z} \right) \in \mathcal{S}_{\Phi_{B}}} w\left(\vec{z}\right),
			\tag{B2} \label{eqB2}
		\end{align}
		where the first equality holds due to Eq. (\ref{eq4}), the
		penultimate one comes from the orthogonality relation $\text{Tr}
		\left[w\left(\vec{y} \right) w\left(\vec{x} \right)\right]=d^{n_{A}}
		\delta_{\vec{x},-\vec{y}}$ and the last one holds by letting
		\[
		\mathcal{S}_{\Phi_{B}}:=\left\lbrace w\left(\vec{z}\right): \exists
		\,\ w\left(\vec{y}\right)^{\dagger} \in \mathcal{S}_{\rho^{T} } \,\
		s.t. \,\  w\left(\vec{y}\right) \otimes w\left(\vec{z}\right) \in
		\mathcal{S}_{J_{\Phi}} \right\rbrace.
		\]
		
		We now show that $\mathcal{S}_{\Phi_{B}}$ is an abelian subgroup of
		$\mathcal{P}$. We can easily verify that $\mathcal{S}_{\Phi_{B}}$ is
		a subgroup of $\mathcal{P}$ because $\mathcal{S}_{J_{\Phi}}$ and
		$\mathcal{S}_{\rho^{T}}$ are subgroups of $\mathcal{P}$. Thus, we
		need to verify the commutativity of $\mathcal{S}_{\Phi_{B}}$.
		
		For any $w\left(\vec{z}_{1}\right), w\left(\vec{z}_{2}\right) \in S_{\Phi_{B}}$, there exist $w\left(\vec{y}_{1}\right)^{\dagger}, w\left(\vec{y}_{2}\right)^{\dagger} \in \mathcal{S}_{\rho^{T}}$ such that
		\[
		w\left(\vec{y}_{i}\right) \otimes w\left(\vec{z}_{i}\right) \in \mathcal{S}_{J_{\Phi}}, \,\ i=1,2,
		\]
		thus we obtain that
		\begin{align*}
			& \left[w\left(\vec{y}_{1}\right),w\left(\vec{y}_{2}\right) \right] \otimes w\left(\vec{z}_{1}\right)w\left(\vec{z}_{2}\right)+w\left(\vec{y}_{2}\right)w\left(\vec{y}_{1}\right) \otimes \left[w\left(\vec{z}_{1}\right),w\left(\vec{z}_{2}\right) \right]\\
			=&  \left[w\left(\vec{y}_{1}\right) \otimes w\left(\vec{z}_{1}\right), w\left(\vec{y}_{2}\right) \otimes w\left(\vec{z}_{2}\right) \right]\\
			=& 0,
		\end{align*}
		where the first equality follows from the fact that $\mathcal{S}_{J_{\Phi}}$ is abelian. Since $\mathcal{S}_{\rho^{T}}$ is also an abelian group, we have
		\[
		w\left(\vec{y}_{2}\right)w\left(\vec{y}_{1}\right) \otimes \left[w\left(\vec{z}_{1}\right),w\left(\vec{z}_{2}\right) \right]=0,
		\]
		which implies that
		\begin{align*}
			\left[w\left(\vec{z}_{1}\right),w\left(\vec{z}_{2}\right) \right]=0.
		\end{align*}
		In summary, $\mathcal{S}_{\Phi_{B}}$ is an abelian subgroup of
		$\mathcal{P}$. Thus from Eq. (\ref{eqB2}), we have $\Phi \left(\rho\right) \in \mathcal{MSPS}$,
		which implies that Eq. (\ref{eqB1}) holds.
		
		Proceeding to the second part, we show that $\Phi \in \mathcal{CMPO}$. Given an ancillary system $R$ with arbitrary dimension, the Choi state of $\text{id}_{R} \otimes \Phi$ is
		\[
		J_{\text{id}_{R} \otimes \Phi}=J_{\text{id}_{R}} \otimes J_{\Phi} \in \mathcal{MSPS},
		\]
		since $J_{\text{id}_{R}}, J_{\Phi} \in \mathcal{MSPS}$ and the tensor product of two MSPSs is still an MSPS. Thus based on Eq. (\ref{eqB1}), we obtain that
		\[
		\left(\text{id}_{R} \otimes \Phi\right) \left(\rho\right) \in \mathcal{MSPS},
		\]
		which implies that $\Phi \in \mathcal{CMPO}$ when $J_{\Phi} \in \mathcal{MSPS}$.
		
		For sufficiency, if $\Phi \in \mathcal{CMPO}$, we have
		\[
		J_{\Phi}=\left(\text{id}_{A} \otimes \Phi\right) \left(|\Omega^{AA^{\prime}} \rangle \langle \Omega^{AA^{\prime}}|\right) \in \mathcal{MSPS}
		\]
		due to the fact that $|\Omega^{AA^{\prime}} \rangle \langle
		\Omega^{AA^{\prime}}|\in \mathcal{MSPS}$.
		
		Hence we complete the proof of Theorem \ref{thm1}. $\hfill\qedsymbol$
		
		\refstepcounter{subsection}
		\subsection*{Appendix \thesubsection. Proof of Proposition \ref{pro2}}\label{Appendix C}

		We first prove that $\mathcal{MPO} \subsetneq \mathcal{SPO}$. The proof for $\mathcal{CMPO} \subsetneq \mathcal{CSPO}$ follows identically due to the linearity of $\operatorname{id}_{R} \otimes \Phi$ for any ancillary system $R$.
		
		Suppose that $\rho \in \mathcal{STAB}$ and $\Phi \in \mathcal{MPO}$. Then we have
		\[
		\Phi\left(\rho\right)=\Phi\left(\sum_{i} p_{i} \psi_{i}\right)=\sum_{i} p_{i} \Phi \left(\psi_{i}\right),
		\]
		where $\left\lbrace \psi_{i} \right\rbrace _{i} \subseteq \mathcal{PS}$ and $\left\lbrace p_{i} \right\rbrace_{i} $ is a classical probability distribution.
		
		Since $\mathcal{PS} \subsetneq \mathcal{MSPS}$, we obtain that
		\[
		\Phi \left(\psi_{i}\right) \in \mathcal{MSPS} \subsetneq \mathcal{STAB},
		\]
		which indicates that $\Phi \left(\rho\right) \in  \mathcal{STAB}$. Furthermore, for the depolarizing channel $\mathcal{D}_p$ with $0 <p <1$ which acts as \cite{Nielsen}
		\[
		\mathcal{D}_p(\rho) = (1-p)\rho + p\frac{\mathbb{I}_d}{d},
		\]
		one can verify that $\mathcal{D}_p \in \mathcal{SPO}$ but $\mathcal{D}_p \notin \mathcal{MPO}$. Thus, we obtain that $\mathcal{MPO} \subsetneq \mathcal{SPO}$.
		
		Moreover, the proof of $\mathcal{CSPO} \subsetneq \mathcal{SPO}$ is trivial, and thus the proof is complete. $\hfill\qedsymbol$
		
		\refstepcounter{subsection}
		\subsection*{Appendix \thesubsection. Proof of Theorem \ref{thm2}}\label{Appendix D}

		What we need to prove is that $\Xi_{\mathcal{S},i}\left(\Phi\right)$ and $\Xi_{\mathcal{S},5}\left(\Theta\right)$ are free channels, $i=1,2,3,4$, where $\Phi \in \mathcal{O}_{A \rightarrow B}$ and $\Theta \in \mathcal{O}_{A \rightarrow BR}$ are free channels.
		
		We only show the cases $i=1,3,5$. The cases $i=2$ and $i=4$ are analogous to the proofs of the cases $i=1$ and $i=3$, respectively.
		
		Suppose that $\mathcal{E} \in \mathcal{CMPO}$ and $\mathcal{E}_{U}$ is a Clifford
		channel. Noting that by Theorem \ref{thm1},
		\begin{align}
			\Xi_{\mathcal{S},1}\left(\Phi\right), \Xi_{\mathcal{S},3}\left(\Phi\right), \Xi_{\mathcal{S},5}\left(\Theta\right) \in \mathcal{CMPO}
			\tag{D1} \label{eqD1}
		\end{align}
		are equivalent to
		\begin{align}
			J_{\mathcal{E} \otimes \Phi}, J_{\mathcal{E}_{U} \circ \Phi}, J_{\mathcal{T}_{R} \circ \Theta} \in \mathcal{MSPS},
			\tag{D2} \label{eqD2}
		\end{align}
		respectively, it suffices to prove Eq. (\ref{eqD2}).
		
		The case $i=1$ is obvious by using $J_{\mathcal{E} \otimes \Phi}=J_{\mathcal{E}} \otimes J_{\Phi}$ and the closedness of MSPSs under tensor product.
		
		As for the case $i=3$, what we need to show is the closedness of MSPSs under Clifford channels, which follows from the fact that
		\[
		J_{\mathcal{E}_{U} \circ \Phi }= \left(\mathbb{I}_{A} \otimes U\right) J_{\Phi} \left(\mathbb{I}_{A} \otimes U\right)^{\dagger}.
		\]
		Suppose that the quantum state $\rho$ on $\mathcal{H}^{\otimes n}$ is a minimal stabilizer-projection state which corresponds to the stabilizer group $\mathcal{S}_{\rho}$. Then we obtain that
		\[
		\mathcal{E}_{U}\left(\rho\right)=U\rho U^{\dagger}=U \left(\frac{1}{d^{n}} \sum_{g \in \mathcal{S}_{\rho}}^{} g\right) U^{\dagger} =\frac{1}{d^{n}} \sum_{g \in \mathcal{S}_{\rho}}^{} UgU^{\dagger} \in \mathcal{MSPS},
		\]
		for any Clifford unitary $U$, where the first two equalities come from the definitions and the membership relation holds as $\left\lbrace UgU^{\dagger}\right\rbrace_{g \in \mathcal{S}_{\rho}} $ is still an abelian subgroup of $\mathcal{P}$.
		
		For the case $i=5$, we omit the phase factors of Weyl operators which have no impact on the result, as in the proof of Theorem \ref{thm1}. Since $\Theta \in \mathcal{CMPO}$, we suppose that
		\[
		J_{\Theta}=\frac{1}{d^{n_{A}+n_{B}+n_{R}}} \sum_{w\left(\vec{x}\right)  \in \mathcal{S}_{J_{\Theta}}} w\left(\vec{x}\right) \in \mathcal{MSPS},
		\]
		where $\mathcal{S}_{J_{\Theta}}$ is the stabilizer group of $J_{\Theta}$. Thus, it holds that
		\begin{align*}
			& J_{\mathcal{T}_{R} \circ \Theta}\\
			=& \left( \text{id}_{AB} \otimes \text{Tr}_{R}\right) \left[J_{\Theta} \right] \\
			=& \frac{1}{d^{n_{A}+n_{B}+n_{R}}}  \sum_{w\left(\vec{y} \right) \otimes w\left(\vec{z}\right)\in \mathcal{S}_{J_{\Theta}}} w\left(\vec{y}\right) \otimes \text{Tr} \left[ w\left(\vec{z}\right)\right] \\
			=&  \frac{1}{d^{n_{A}+n_{B}}} \sum_{w\left(\vec{y}\right) \in \mathcal{S}_{AB}} w\left(
			\vec{y}\right) ,
		\end{align*}
		where the second equality follows from the linearity of taking partial trace and the third one comes from the fact that $\text{Tr}\left[w\left(\vec{z}\right)\right]=d^{n_{R}} \delta_{\vec{z},\vec{0}}$.
		
		Furthermore, based on the fact that $\mathcal{S}_{AB}=\left\lbrace
		w\left(\vec{y}\right): w\left(\vec{y}\right) \otimes \mathbb{I}_{R}
		\in \mathcal{S}_{J_{\Theta}}  \right\rbrace $ is an abelian subgroup
		of $\mathcal{P}$,  we obtain that $J_{\mathcal{T}_{R} \circ \Theta}
		\in \mathcal{MSPS}$. This completes the proof. $\hfill \qedsymbol$
		
		\refstepcounter{subsection}
		\subsection*{Appendix \thesubsection. Proof of Lemma \ref{lemma1}}\label{Appendix E}
		
		Item (i) directly follows from Theorem \ref{thm1} and Proposition 1 in \cite{Bu1}.
		
		For item (ii), it holds that
		\begin{align}
			& \mathcal{D}_{Rz}\left(\Phi_{1} \otimes \Phi_{2} \middle\| \Theta_{1} \otimes \Theta_{2} \right) \notag\\
			=& D_{Rz} \left(J_{\Phi_{1} \otimes \Phi_{2}} \middle\| J_{\Theta_{1} \otimes \Theta_{2}}\right)\notag\\
			=& D_{Rz} \left(J_{\Phi_{1} }\otimes J_{\Phi_{2}} \middle\| J_{\Theta_{1}} \otimes J_{\Theta_{2}}\right)\notag\\
			=& D_{Rz} \left(J_{\Phi_{1}} \middle\| J_{\Theta_{1}}\right)+D_{Rz} \left(J_{\Phi_{2}} \middle\| J_{\Theta_{2}}\right)\notag\\
			=& \mathcal{D}_{Rz} \left(\Phi_{1} \middle\| \Theta_{1} \right)+\mathcal{D}_{Rz} \left(\Phi_{2} \middle\| \Theta_{2}\right),
			\tag{E1} \label{eqE1}
		\end{align}
		where the second equality comes from the fact that $J_{\Phi \otimes
			\Theta}= J_{\Phi} \otimes J_{\Theta}$, the penultimate follows from
		Proposition 1 in \cite{Bu1} and the others follow from the
		definitions.
		
		To prove item (iii), observe that
		\begin{align}
			& \mathcal{D}_{Rz} \left(\mathcal{E}_{U} \circ \Phi \middle\| \mathcal{E}_{U} \circ \Theta \right)\notag\\
			=& D_{Rz} \left(J_{\mathcal{E}_{U} \circ \Phi  } \middle\| J_{\mathcal{E}_{U} \circ \Theta } \right)\notag\\
			=& D_{Rz} \left( \left(\mathbb{I}_{A} \otimes U\right) J_{\Phi} \left(\mathbb{I}_{A} \otimes U\right)^{\dagger}   \middle\| \left(\mathbb{I}_{A} \otimes U\right) J_{\Theta} \left(\mathbb{I}_{A} \otimes U\right)^{\dagger}\right)\notag\\
			=& D_{Rz} \left(J_{\Phi} \middle\| J_{\Theta}\right)\notag\\
			=& \mathcal{D}_{Rz} \left(\Phi \middle\| \Theta \right),
			\tag{E2} \label{eqE2}
		\end{align}
		where the second equality comes from the fact that for any Clifford
		channel $\mathcal{E}_{U}$, we have $	J_{\mathcal{E}_{U} \circ \Phi }= \left(\mathbb{I}_{A} \otimes U\right) J_{\Phi} \left(\mathbb{I}_{A} \otimes U\right)^{\dagger}$,	the penultimate one follows from the evidence that $\mathbb{I}_{A}
		\otimes U$ is still a Clifford unitary and Proposition 1 in
		\cite{Bu1}, and the others come from the definitions.
		
		Using the fact that $J_{\Theta \circ \mathcal{E}_{V} }=  \left(V^{T} \otimes \mathbb{I}_{B}\right)  J_{\Theta} \left(V^{T} \otimes  \mathbb{I}_{B}\right) ^{\dagger}$ and the closedness of Clifford unitaries under transpose, we can verify that $\mathcal{D}_{Rz} \left(\Phi \circ \mathcal{E}_{V}  \middle\| \Theta \circ \mathcal{E}_{V} \right)=\mathcal{D}_{Rz} \left(\Phi \middle\| \Theta\right)$. So item (iii) holds.
		
		For item (iv), we present the proof for the convexity in the first argument, and the concavity in the second argument follows an identical line of reasoning. For the former one, we have
		\begin{align}
			&\mathcal{D}_{Rz} \left(\sum_{i} p_{i} \Phi_{i} \middle\| \Theta \right) \notag\\
			=& D_{Rz} \left(J_{\sum_{i} p_{i} \Phi_{i}} \middle\| J_{\Theta}\right)\notag\\
			=& D_{Rz} \left(\sum_{i} p_{i} J_{\Phi_{i}} \Big\| J_{\Theta}\right)\notag\\
			\leq & \sum_{i} p_{i} D_{Rz} \left(J_{\Phi_{i}} \middle\| J_{\Theta}\right)\notag\\
			=& \sum_{i} p_{i} \mathcal{D}_{Rz} \left(\Phi_{i}\middle\|\Theta\right),
			\tag{E3} \label{eqE3}
		\end{align}
		where the second equality comes from the fact that $J_{\sum_{i}
			p_{i} \Phi_{i}}=\sum_{i} p_{i} J_{\Phi_{i}}$, and the inequality
		comes from Proposition 1 in \cite{Bu1}.
		
		In summary, combining Eqs. (\ref{eqE1})-(\ref{eqE3}), we complete the proof of Lemma \ref{lemma1}.$\hfill\qedsymbol$
		
		\refstepcounter{subsection}
		\subsection*{Appendix \thesubsection. Proof of Theorem \ref{thm3}}\label{Appendix F}
		
		To prove Theorem \ref{thm3}, it suffices to show that $\mathcal{M}_{Rz}$ is monotonically non-increasing under each elementary free superoperation defined in Definition \ref{def2}.
		
		Firstly, given any Clifford channels $\mathcal{E}_{U}, \mathcal{E}_{V} $, it holds that
		\begin{align}
			&\mathcal{M}_{Rz} \left(\mathcal{E}_{U} \circ \Phi \right)\notag\\
			=& \min_{\Theta \in \mathcal{CMPO}} \mathcal{D}_{Rz} \left(\mathcal{E}_{U} \circ \Phi \middle\| \Theta\right)\notag\\
			= & \min_{\Theta \in \mathcal{CMPO}} \mathcal{D}_{Rz} \left(\mathcal{E}_{U} \circ \Phi \middle\| \mathcal{E}_{U} \circ \Theta  \right)\notag\\
			= & \min_{\Theta \in \mathcal{CMPO}} \mathcal{D}_{Rz} \left(\Phi \middle\| \Theta\right)\notag\\
			=& \mathcal{M}_{Rz} \left(\Phi\right),
			\tag{F1}\label{eqF1}
		\end{align}
		where the second equality comes from the fact that $\mathcal{E}_{U} \circ \Theta  \in \mathcal{CMPO}$ iff $\mathcal{E}_{U}$ is a Clifford channel when $\Theta \in \mathcal{CMPO}$, the third one comes from Lemma \ref{lemma1} (iii) and the others follow from the definitions.
		
		The proof of  $\mathcal{M}_{Rz} \left(\Phi \circ \mathcal{E}_{V} \right)=\mathcal{M}_{Rz} \left(\Phi\right)$ is similar to the above one, thus we omit it.
		
		Secondly, suppose that $\mathcal{E}^{*}$ and $\mathcal{E}^{**}$
		attain the minimal values of $\mathcal{M}_{Rz}\left(\Phi\right)$ and
		$\mathcal{M}_{Rz} \left(\Theta\right)$, respectively, i.e.,
		\[
		\mathcal{M}_{Rz}\left(\Phi\right)=\mathcal{D}_{Rz}\left(\Phi \middle\| \mathcal{E}^{*}\right) \,\ \text{and} \,\ \mathcal{M}_{Rz}\left(\Theta\right)=\mathcal{D}_{Rz}\left(\Theta \middle\| \mathcal{E}^{**}\right),
		\]
		where $\mathcal{E}^{*}, \mathcal{E}^{**} \in \mathcal{CMPO}$. Thus we have
		\begin{align*}
			&\mathcal{M}_{Rz}\left(\Phi \otimes \Theta\right)\\
			=& \min_{ \mathcal{E} \in \mathcal{CMPO} } \mathcal{D}_{Rz} \left(\Phi \otimes \Theta \middle\| \mathcal{E}\right)\\
			\leq &  \mathcal{D}_{Rz} \left(\Phi \otimes \Theta \middle\| \mathcal{E}^{*} \otimes \mathcal{E}^{**}\right)\\
			=&  \mathcal{D}_{Rz}\left(\Phi \middle\| \mathcal{E}^{*}\right)+ \mathcal{D}_{Rz}\left(\Theta \middle\| \mathcal{E}^{**}\right)\\
			=&\mathcal{M}_{Rz}\left(\Phi\right)+\mathcal{M}_{Rz}\left(\Theta\right),
		\end{align*}
		where the first inequality comes from the fact that $\mathcal{E}^{*}\otimes \mathcal{E}^{**} \in \mathcal{CMPO}$ since both $\mathcal{E}^{*}, \mathcal{E}^{**}$ are free channels, and the penultimate one comes from Lemma \ref{lemma1} (ii). Since $\mathcal{M}_{Rz}\left(\mathcal{E}\right)=0$ whenever $\mathcal{E} \in \mathcal{CMPO}$, we have
		\begin{align}
			\mathcal{M}_{Rz} \left(\Phi \otimes \mathcal{E}_{1} \right) \leq \mathcal{M}_{Rz} \left(\Phi\right),
			\tag{F2} \label{eqF2}
		\end{align}
		and
		\begin{align}
			\mathcal{M}_{Rz} \left(\mathcal{E}_{2} \otimes \Phi\right)  \leq \mathcal{M}_{Rz} \left(\Phi\right),
			\tag{F3} \label{eqF3}
		\end{align}
		where $\mathcal{E}_{1},\mathcal{E}_{2} \in \mathcal{CMPO}$.
		
		Finally, it holds that
		\begin{align}
			& \mathcal{M}_{Rz}\left(\mathcal{T}_{R} \circ \Theta\right)\notag\\
			=& \min_{\mathcal{E} \in \mathcal{CMPO}} \mathcal{D}_{Rz} \left(\mathcal{T}_{R} \circ \Theta \middle\| \mathcal{E}\right)\notag\\
			=& \min_{\mathcal{E} \in \mathcal{CMPO}} D_{Rz} \left(J_{\mathcal{T}_{R} \circ \Theta }\middle\| J_{\mathcal{E}}\right)\notag\\
			\le & \min_{\mathcal{E}^{\prime} \in \mathcal{CMPO}} D_{Rz} \left(\text{Tr}_{R}\left[J_{\Theta}\right]  \middle\| \text{Tr}_{R}\left[J_{\mathcal{E}^{\prime}}\right] \right)\notag\\
			\le &  \min_{\mathcal{E}^{\prime} \in \mathcal{CMPO}} D_{Rz} \left(J_{\Theta} \middle\| J_{\mathcal{E}^{\prime}}\right) \notag\\
			=& \mathcal{M}_{Rz}\left(\Theta\right),
			\tag{F4} \label{eqF4}
		\end{align}
		where the second equality holds because $J_{\mathcal{T}_{R} \circ
			\Theta}=\text{Tr}_{R}\left[J_{\Theta}\right]$, the first inequality follows from the fact that $\mathcal{T}_{R} \circ \mathcal{E}^{\prime} \in \mathcal{CMPO}$ if $\mathcal{E}^{\prime} \in \mathcal{CMPO}$ and the second
		one follows from Proposition 1 of \cite{Bu1}.
		
		Thus, combining Eqs. (\ref{eqF1})-(\ref{eqF4}), we complete the proof of Theorem \ref{thm3}.
		$\hfill\qedsymbol$

		\refstepcounter{subsection}
		\subsection*{Appendix \thesubsection. Proof of Proposition \ref{pro3}}\label{Appendix G}

		From Eqs. (\ref{eq9}) and (\ref{eq10}), there exists a pure stabilizer state $\psi$ such that
		\begin{align}
			\mathcal{M}_{Rz}\left(\Phi\right)=\min_{\sigma \in  \mathcal{MSPS}, \text{Tr}_{B} \sigma=\mathbb{I}_{A}/d^{n_A}} D_{Rz}\left(J_{\Phi} \middle\| \sigma\right) \geq \min_{\sigma \in  \mathcal{MSPS}} D_{Rz}\left(J_{\Phi} \middle\| \sigma\right) =S\left(J_{\Phi} \boxtimes \psi\right)-S\left(J_{\Phi}\right).
			\tag{G1} \label{eqG1}
		\end{align}
		Furthermore, the stabilizer group $\mathcal{S}_{\psi}$ of $\psi$ yields an orthonormal basis $B_{\mathcal{S}_{\psi}}=\left\lbrace |\psi_{\vec{k}} \rangle\right\rbrace_{\vec{k} \in \mathbb{Z}_{d}^{n_{A}+n_{B}}} $.
		
		Assume that $\Delta_{B_{\mathcal{S}_{\psi}}}$ is the fully-dephasing channel with respect to the basis $B_{\mathcal{S}_{\psi}}$ acting as
		\[
		\Delta_{B_{\mathcal{S}_{\psi}}}\left(J_{\Phi}\right)=\sum_{\vec{k}} \langle \psi_{\vec{k}}| J_{\Phi} | \psi_{\vec{k}}\rangle | \psi_{\vec{k}}\rangle \langle \psi_{\vec{k}}| ,
		\]
		thus we have
		\begin{align}
			&D\left(J_{\Phi}  \middle\|  \Delta_{B_{\mathcal{S}_{\psi}}}\left(J_{\Phi}\right)\right)\notag\\
			=& -\text{Tr}\left[J_{\Phi} \log \Delta_{B_{\mathcal{S}_{\psi}}}\left(J_{\Phi}\right)\right]-S\left(J_{\Phi}\right)\notag\\
			=& -\sum_{\vec{k}} \langle \psi_{\vec{k}}| J_{\Phi} | \psi_{\vec{k}}\rangle \log \langle \psi_{\vec{k}}| J_{\Phi} | \psi_{\vec{k}}\rangle-S\left(J_{\Phi}\right)\notag\\
			=& S\left(J_{\Phi} \boxtimes \psi \right)-S\left(J_{\Phi}\right)\notag\\
			\le& \mathcal{M}_{Rz}\left(\Phi\right),
			\tag{G2} \label{eqG2}
		\end{align}
		where the second equality holds by a direct calculation, the penultimate one follows from the result established in Proposition 3 of \cite{Bu1} and the inequality follows from Eq. (\ref{eqG1}). Thus we have
		\[
		\mathcal{M}_{R}\left(\Phi\right)=\min_{J_{\Theta} \in \mathcal{STAB}} D\left(J_{\Phi} \middle\| J_{\Theta}\right) \leq D\left(J_{\Phi}  \middle\|  \Delta_{B_{\mathcal{S}_{\psi}}}\left(J_{\Phi}\right)\right)\le \mathcal{M}_{Rz}\left(\Phi\right),
		\]
		where the first inequality holds because
		$\Delta_{B_{\mathcal{S}_{\psi}}}\left(J_{\Phi}\right)$ is a
		stabilizer state and the second inequality comes from Eq. (\ref{eqG2}).  $\hfill \qedsymbol$
		
			\refstepcounter{subsection}
		\subsection*{Appendix \thesubsection. Proof of Theorem \ref{thm5}}\label{Appendix H}
		
		First, we confirm the stabilizer group of $J_{\Phi}$ so that we can obtain $\mathcal{M}(J_\Phi)$. Since \(\Phi\) is trace preserving, we obtain
	\begin{equation}
			\Phi^\dagger(w(\vec y_i))=\zeta_iw(\vec x_i) \Leftrightarrow w(\vec y_i)K_j=\zeta_iK_jw(\vec x_i),\quad\forall j,
				\tag{H1} \label{eqH1}
	\end{equation}
where the sufficiency follows from the definition of the dual channel and the necessity follows from the fact that $\sum_j\left\|w(\vec y_i)K_j-\zeta_iK_jw(\vec x_i)\right\|_2^2=0$.

Define the joint Weyl operators $g_i=\zeta_i^{-1}\left(\overline{w(\vec x_i)}\otimes w(\vec y_i)\right)$, where  $\overline{w(\vec x_i)}$ denotes the complex conjugate of $w(\vec x_i)$. Then using the vectorization identity \cite{Watrous}
\begin{equation}
	|AFB\rangle\!\rangle=(B^T\otimes A)|F\rangle\!\rangle,
	\tag{H2} \label{eqH2}
\end{equation} 
we obtain
\[
	g_i|K_j\rangle\!\rangle=|K_j\rangle\!\rangle,
\qquad
|\mathcal C_i(F)\rangle\!\rangle=g_i|F\rangle\!\rangle,
\]
for every linear operator $F$, where \(|F\rangle\!\rangle\) denotes the vectorization of \(F\). 

Suppose the stabilizer group of \(J_\Phi\) is $S_{J_\Phi}$, now we show \(S_{J_\Phi}=\langle g_1,\ldots,g_r\rangle\). By the Choi–Kraus representation, we have
\begin{equation}
	J_\Phi=\frac1{d^{n_A}}\sum_j|K_j\rangle\!\rangle\langle\!\langle K_j|.
		\tag{H3} \label{eqH3}
\end{equation}
 Then consider an arbitrary operator of the form $g=\zeta^{-1}\bigl(\overline{w(\vec x)}\otimes w(\vec y)\bigr)$, where $\zeta=\omega^{k}$ with $ k \in \mathbb{Z}_{d}$. Therefore, from Eq. (\ref{eqH1}) to Eq. (\ref{eqH3}), it holds that
\begin{equation}
	gJ_\Phi=J_\Phi
	\iff
	g|K_j\rangle\!\rangle=|K_j\rangle\!\rangle,\quad\forall j
	\iff
	w(\vec y)K_j=\zeta K_jw(\vec x),\quad\forall j
	\iff
	\Phi^\dagger(w(\vec y))=\zeta w(\vec x).
		\tag{H4} \label{eqH4}
\end{equation}
 Since the chosen \(r\) relations form a complete set of independent generators, it holds that $ S_{J_\Phi}
 \subseteq\langle g_1,\ldots,g_r\rangle$. Moreover, from Eq. (\ref{eqH4}), we have $ \langle g_1,\ldots,g_r\rangle \subseteq S_{J_\Phi}$. Therefore, we obtain
 \[
 S_{J_\Phi}=\langle g_1,\ldots,g_r \rangle.
 \]
 Then by the definition of the mean state, we have
\begin{equation}
	\mathcal{M}(J_\Phi)=\frac1{d^{n_{A}+n_{B}}}\sum_{g\in S_{J_\Phi}}g.
		\tag{H5} \label{eqH5}
\end{equation}

Now we construct the Kraus operators of \(\widetilde{\mathcal{M}}(\Phi)\). Since the Pauli operators \(g_i\) share a common eigenvector with eigenvalue $+1$, then they commute. Moreover, \(g_i^d=\mathbb{I}\). Consequently, \(\mathcal P_\Phi\) is an orthogonal projection satisfying
\begin{equation}
		|\mathcal P_\Phi(F)\rangle\!\rangle
	=
	\left(\frac1{d^r}\sum_{g\in S_{J_\Phi}}g\right)
	|F\rangle\!\rangle.
		\tag{H6} \label{eqH6}
\end{equation}
Let \(E_{\vec b,\vec a}=|\vec b\rangle\langle\vec a|\), where \(\vec a\in\mathbb Z_d^{n_A}\) and \(\vec b\in\mathbb Z_d^{n_B}\). Since $\left\lbrace E_{\vec b,\vec a} \right\rbrace_{\vec{b},\vec{a}}$ forms a Hilbert–Schmidt orthonormal basis, it holds that
\begin{equation}
	\sum_{\vec a,\vec b}
	|E_{\vec b,\vec a}\rangle\!\rangle
	\langle\!\langle E_{\vec b,\vec a}|=\mathbb{I}_{AB}.
	\tag{H7} \label{eqH7}
\end{equation}
Using Eqs. (\ref{eqH6}) and (\ref{eqH7}), we obtain
\[
\sum_{\vec a,\vec b} |\mathcal P_\Phi(E_{\vec b,\vec a})\rangle\!\rangle \langle\!\langle\mathcal P_\Phi(E_{\vec b,\vec a})| =\frac1{d^r}\sum_{g\in S_{J_\Phi}} g.
\]
Consequently, we have
\begin{equation}
	\mathcal{M} (J_\Phi)=\frac{1}{d^{n_{A}+n_{B}-r}} \sum_{\vec a,\vec b} |\mathcal P_\Phi(E_{\vec b,\vec a})\rangle\!\rangle\langle\!\langle\mathcal P_\Phi(E_{\vec b,\vec a})|.
		\tag{H8} \label{eqH8}
\end{equation}
Comparing Eqs. (\ref{eqH3}) and (\ref{eqH8}), we identify
\[
\widetilde K_{\vec b,\vec a}
=
\sqrt{d^{r-n_{B}}}\,
\mathcal P_\Phi(E_{\vec b,\vec a}).
\]

Furthermore, since \(\Phi\) is trace preserving and any \(g_A\otimes \mathbb{I}_B\in S_{J_\Phi}\) must satisfy \(g_A=\mathbb{I}_A\). Thus,
\[
\operatorname{Tr}_B \mathcal{M}(J_\Phi)=\frac{\mathbb{I}_A}{d^{n_A}},
\]
which implies that
\[
\sum_{\vec a,\vec b}
\widetilde K_{\vec b,\vec a}^\dagger
\widetilde K_{\vec b,\vec a}=\mathbb{I}_A.
\]

Therefore, the constructed Kraus operators define a CPTP map whose Choi state is \(\mathcal{M}(J_\Phi)\). By the definition of the mean channel, this map is precisely \(\widetilde{\mathcal{M}}(\Phi)\). This completes the proof. $\hfill\qedsymbol$

		\refstepcounter{subsection}
		\subsection*{Appendix \thesubsection. Proof of Theorem \ref{thm6}}\label{Appendix I}
		
	For any zero-mean channel $\Phi$, $J_\Phi$ is a zero-mean state.
	By Lemma 90 in \cite{Bu3}, it can be proved inductively that $\mathcal{M}(\boxtimes^N J_\Phi)
	=\boxtimes^N\mathcal{M}(J_\Phi).$
	Moreover, since $J_\Phi$ is a zero-mean state, it follows from
	the convolution-multiplication duality and the group structure
	of $S_\Phi$ that $	\Xi_{\boxtimes^N J_\Phi}(\vec{x})=1$ iff $\Xi_{J_\Phi}(\vec{x})=1$.
	Consequently, we have
	\begin{equation}
		\mathcal{M}(J_\Phi)
		=\mathcal{M}(\boxtimes^N J_\Phi)
		=\boxtimes^N\mathcal{M}(J_\Phi),
		\tag{I1} \label{eqI1}
	\end{equation}
	which also implies that $\boxtimes^N J_\Phi$ is a zero-mean
	state. Thus, it holds that
		\begin{align}
			&   \mathcal{D}_{Rz}\left(\boxtimes ^{N} \Phi \middle\| \widetilde{\mathcal{M}}\left(\Phi\right)\right)\notag\\
			=& S\left(J_{\boxtimes ^{N} \Phi} \boxtimes J_{\widetilde{\mathcal{M}}\left(\Phi\right)}\right)-S\left(J_{\boxtimes ^{N} \Phi}\right)\notag\\
			=& S\left(\boxtimes ^{N} J_{\Phi} \boxtimes \mathcal{M}\left(J_{\Phi}\right)\right)-S\left(\boxtimes ^{N} J_{\Phi}\right)\notag\\
			=&  S\left(\boxtimes ^{N} J_{\Phi} \boxtimes \mathcal{M}\left(\boxtimes^{N} J_{\Phi}\right)\right)-S\left(\boxtimes ^{N} J_{\Phi}\right)\notag\\
			=& S\left(\mathcal{M}\left(\boxtimes^{2N} J_{\Phi}\right)\right)-S\left(\boxtimes ^{N} J_{\Phi}\right)\notag\\
			=&  S\left(\mathcal{M}\left( J_{\Phi}\right)\right)-S\left(\boxtimes ^{N} J_{\Phi}\right),
			\tag{I2} \label{eqI2}
		\end{align}
		where the first two equalities come from the definitions, the third and last ones follow from Eq. (\ref{eqI1}) and the penultimate one follows from Lemma 90 in \cite{Bu3}, that is,
		\[
		\boxtimes ^{N} J_{\Phi} \boxtimes \mathcal{M}\left(\boxtimes^{N} J_{\Phi}\right)=\mathcal{M}\left(\boxtimes^{2N} J_{\Phi}\right).
		\]
		
		Subsequently, assume that the corresponding stabilizer group of $\mathcal{M}\left( J_{\Phi}\right)$ is $\mathcal{S}_{\Phi}$ of order $d^{r}$. Then we have $S\left(\mathcal{M}\left( J_{\Phi}\right)\right)=\log d^{2n-r}$ by a direct calculation. Denote $c_0 = s^N, c_k = s^{N-k}t, 1 \le k \le N$. Then it holds that
		\begin{align*}
			&   \mathcal{D}_{Rz}\left(\boxtimes ^{N} \Phi \middle\| \widetilde{\mathcal{M}}\left(\Phi\right)\right)\\
			\leq & \log d^{2n-r}-S_{2}\left(\boxtimes ^{N} J_{\Phi}\right)\\
			=& \log d^{2n-r}+\log\left\lbrace \frac{1}{d^{2n-r}}+ \frac{1}{d^{2n}}\sum_{\vec{x}: w\left(\vec{x}\right) \notin \mathcal{S}_{\Phi}} \left|\Xi_{\boxtimes^N J_{\Phi}}\left(\vec{x}\right) \right|^{2} \right\rbrace \\
			=&  \log d^{2n-r}+\log\left\lbrace \frac{1}{d^{2n-r}}+ \frac{1}{d^{2n}} \sum_{\vec{x}: w\left(\vec{x}\right) \notin \mathcal{S}_{\Phi}} \prod_{k=0}^N \vert{}\Xi_{J_\Phi}(c_k \vec{x})\vert{}^2 \right\rbrace\\
			\leq &  \log d^{2n-r}+\log\left\lbrace \frac{1}{d^{2n-r}}+\frac{1}{d^{2n}}  \prod_{k=0}^N \left(  \sum_{\vec{x}: w\left(\vec{x}\right) \notin \mathcal{S}_{\Phi}} \vert{}\Xi_{J_\Phi}(c_k \vec{x})\vert{}^{2N+2} \right)^{\frac{1}{N+1}} \right\rbrace\\
			=& \log \left\lbrace 1+ \frac{1}{d^r} \sum_{\vec{x}: w\left(\vec{x}\right) \notin \mathcal{S}_{\Phi}} \vert{}\Xi_{J_\Phi}(\vec{x})\vert{}^{2N+2} \right\rbrace,
		\end{align*}
		where the first inequality comes from the monotonicity of R\'{e}nyi entropy and Eq. (\ref{eqI2}), the first equality follows from the fact that
		\[
		-S_{2}\left(\boxtimes ^{N} J_{\Phi}\right)=\log \left\lbrace \text{Tr}\left[\left( \boxtimes ^{N} J_{\Phi}\right) ^{2}\right]\right\rbrace =\log\left\lbrace \frac{1}{d^{2n-r}}+ \frac{1}{d^{2n}}\sum_{\vec{x}: w\left(\vec{x}\right) \notin \mathcal{S}_{\Phi}} \left|\Xi_{\boxtimes^N J_{\Phi}}\left(\vec{x}\right) \right|^{2} \right\rbrace,
		\]
		the second one comes from the fact that
		\[
		\Xi_{\boxtimes^N J_\Phi}(\vec{x}) = \Xi_{J_\Phi}(s^N \vec{x})\Xi_{J_\Phi}(s^{N-1}t \vec{x}) \cdots \Xi_{J_\Phi}(st \vec{x})\Xi_{J_\Phi}(t \vec{x}),
		\]
		the second inequality follows from the generalized discrete Hölder's inequality for $N+1$ functions, with conjugate exponents $p_0 = p_1 = \dots = p_N = N+1$, that is 
		\[
		\sum_{\vec{x}: w\left(\vec{x}\right) \notin \mathcal{S}_{\Phi}} \prod_{k=0}^N \vert{}\Xi_{J_\Phi}(c_k \vec{x})\vert{}^2 \leq  \prod_{k=0}^N \left(  \sum_{\vec{x}: w\left(\vec{x}\right) \notin \mathcal{S}_{\Phi}} \vert{}\Xi_{J_\Phi}(c_k \vec{x})\vert{}^{2N+2} \right)^{\frac{1}{N+1}},
		\]
		and the last equality follows from the fact that $c_k \neq 0$ in $\mathbb{Z}_d$, thus, for any $\vec{x}$, it holds that
		\[
		w\left(\vec{x}\right) \notin \mathcal{S}_{\Phi} \Leftrightarrow w\left(c_k \vec{x}\right) \notin \mathcal{S}_{\Phi}.
		\]
		
		Now consider the following two cases.
		
		(1) If $J_{\Phi} \in \mathcal{MSPS}$, we have $\boxtimes^{N}
		J_{\Phi} \in \mathcal{MSPS}$ from Lemma 50 in \cite{Bu3}. Since
		$\mathcal{M}$ is a resource-destroying map, we get $
		\mathcal{M}\left(\boxtimes^{N} J_{\Phi}\right)=\boxtimes^{N}
		J_{\Phi}$. Moreover, $S\left(\rho \boxtimes
		\rho\right)-S\left(\rho\right)=0$ iff $\rho \in \mathcal{MSPS}$.
		Therefore, it holds that
		\begin{align}
			\mathcal{D}_{Rz}\left(\boxtimes ^{N} \Phi \middle\| \widetilde{\mathcal{M}}\left(\Phi\right)\right)
			=& S\left(\boxtimes ^{N} J_{\Phi} \boxtimes \mathcal{M}\left(\boxtimes^{N} J_{\Phi}\right)\right)-S\left(\boxtimes ^{N} J_{\Phi}\right)\notag\\
			=& S\left(\left[ \boxtimes ^{N} J_{\Phi} \right] \boxtimes \left[ \boxtimes^{N} J_{\Phi}\right] \right)-S\left(\boxtimes ^{N} J_{\Phi}\right)\notag\\
			=& 0.
			\tag{I3} \label{eqI3}
		\end{align}
		
		(2) If $J_{\Phi} \notin \mathcal{MSPS}$, we have $0 \leq \vert{}\Xi_{J_\Phi}(\vec{x})\vert{}
		<1$ for any $\vec{x} \notin \mathcal{S}_{\Phi}$, hence
		\begin{align}
			\lim_{N \rightarrow \infty} \log \left\lbrace 1+ \frac{1}{d^r} \sum_{\vec{x}: w\left(\vec{x}\right) \notin \mathcal{S}_{\Phi}} \vert{}\Xi_{J_\Phi}(\vec{x})\vert{}^{2N+2} \right\rbrace=0.
			\tag{I4} \label{eqI4}
		\end{align}
		In either case, we have $\mathcal{D}_{Rz}\left(\boxtimes ^{N} \Phi
		\middle\| \widetilde{\mathcal{M}}\left(\Phi\right)\right) \rightarrow
		0\,\, (N \rightarrow \infty)$.
		
		Finally, note that for any zero-mean state $\rho$, it follows from Eq. (\ref{eqI1}) and Lemma 90 in \cite{Bu3} that $\rho\boxtimes\mathcal{M}(\rho)=\mathcal{M}(\rho)$.	Hence we have
		\begin{equation}
			D_{Rz}\left( \rho\|\mathcal{M}\left( \rho\right) \right) =S\left( \rho\boxtimes\mathcal{M}\left( \rho\right) \right) -S\left( \rho\right) =S\left( \mathcal{M}\left( \rho\right) \right) -S\left( \rho\right) =D\left( \rho\|\mathcal{M}\left( \rho\right) \right) .
			\tag{I5} \label{eqI5}
		\end{equation}
		This implies that $D_{Rz}\left( \rho\|\mathcal{M}\left( \rho\right) \right) =0$
		iff $\rho\in\mathcal{MSPS}$.
		Hence we claim that $J_{\boxtimes ^{N} \Phi }$ converges to an
		MSPS from Eqs. (\ref{eqI3})-(\ref{eqI5}), which implies that $\boxtimes^{N}
		\Phi$ converges to a completely MSPS-preserving channel when $N
		\rightarrow \infty$. $\hfill\qedsymbol$
		
		\refstepcounter{subsection}
		\subsection*{Appendix \thesubsection. Comparison of Ruzsa divergence-based bound in Eq. (\ref{eq19}) with existing bounds}\label{Appendix J}
		
		First, for any quantum channel $\Phi \in \mathcal{C}_{A \rightarrow B}$, the magic gap of $\Phi$ is defined as the magic gap of $J_{\Phi}$ \cite{Bu3}, i.e., $MG\left(\Phi\right)=MG\left(J_{\Phi}\right)$, and we denote $\lambda=\left(1-MG\left(\Phi\right)\right)^{2}$.
		
		\textbf{(Comparison with the direct diamond norm bound)} For the diamond norm induced Q-CLT for quantum channels, the upper bound is given by \cite{Bu3}
		\begin{align}
			\left\| \boxtimes^N \Phi - \widetilde{\mathcal{M}}(\Phi)\right\| _\diamond \le d^{2n} \lambda^{\frac{N}{2}} \left\| J_\Phi - J_{\widetilde{\mathcal{M}}(\Phi)}\right\| _2=d^n\sqrt{ \sum_{\vec{x}: w\left(\vec{x}\right) \notin \mathcal{S}_{\Phi}} \vert{}\Xi_{J_\Phi}(\vec{x})\vert{}^{2}\lambda^N },
			\tag{J1} \label{eqJ1}
		\end{align}
		where $\left\| \cdot\right\| _\diamond$ denotes the diamond norm and $\left\| \cdot \right\| _2$ is the Schatten $L^2$ norm. Comparing this with our result, it holds that
		\[
		\log \left\lbrace 1+ \frac{1}{d^r} \sum_{\vec{x}: w\left(\vec{x}\right) \notin \mathcal{S}_{\Phi}} \vert{}\Xi_{J_\Phi}(\vec{x})\vert{}^{2N+2}\right\rbrace  \leq \frac{1}{d^r} \sum_{\vec{x}: w\left(\vec{x}\right) \notin \mathcal{S}_{\Phi}} \vert{}\Xi_{J_\Phi}(\vec{x})\vert{}^{2N+2} \leq d^n\sqrt{ \sum_{\vec{x}: w\left(\vec{x}\right) \notin \mathcal{S}_{\Phi}} \vert{}\Xi_{J_\Phi}(\vec{x})\vert{}^{2}\lambda^N },
		\]
		where the second inequality holds for sufficiently large $N$ because the Ruzsa divergence bound decays scaling as $O(\lambda^N)$, which converges to zero exponentially faster than the diamond norm bound that decays scaling as $O(\lambda^{N/2})$, where $f(N) = O(g(N))$ implies that there exists a positive constant $C$ such that $f(N) \le C g(N)$ for sufficiently large $N$. Consequently, the upper bound from the Ruzsa divergence perspective for quantum channels is tighter than the diamond norm version. Furthermore, our bound is entirely independent of the exponentially growing dimension $d^n$ inherent in Eq. (\ref{eqJ1}).
		
		\textbf{(Tighter diamond norm bound via Theorem \ref{thm6})} We can also derive a tighter upper bound for Eq. (\ref{eqJ1}) by applying Theorem \ref{thm6} for $r \geq 1$. Firstly, it holds that
		\begin{align*}
			\left\| \boxtimes^N \Phi - \widetilde{\mathcal{M}}(\Phi)\right\| _\diamond 
			\leq &d^n \left\| J_{\boxtimes^N \Phi} - J_{\widetilde{\mathcal{M}}(\Phi)}\right\|_1 \\
			\leq & d^n \sqrt{2 D\left(  J_{\boxtimes^N \Phi} \big\| J_{\widetilde{\mathcal{M}}(\Phi)} \right)}\\
			=& d^n \sqrt{2 \mathcal{D}_{Rz}\left(\boxtimes ^{N} \Phi \middle\| \widetilde{\mathcal{M}}\left(\Phi\right)\right)},
		\end{align*}
		where the first inequality comes from the fact that $\left\| \Phi_{1}-\Phi_{2}\right\|_{\diamond} \leq d^n \left\| J_{\Phi_{1}}-J_{\Phi_{2}}\right\| _1$, the second one follows from the quantum Pinsker inequality \cite{Watrous}, that is $\vert{}\vert{}\rho - \sigma\vert{}\vert{}_1 \le \sqrt{2 \ln 2 D(\rho \vert{}\vert{} \sigma)}$ and the equality comes from the fact that
		\[
		\mathcal{D}_{Rz}\left(\boxtimes ^{N} \Phi \middle\| \widetilde{\mathcal{M}}\left(\Phi\right)\right)=S\left(\mathcal{M}\left( J_{\Phi}\right)\right)-S\left(\boxtimes ^{N} J_{\Phi}\right)=D\left(  J_{\boxtimes^N \Phi} \big\| J_{\widetilde{\mathcal{M}}(\Phi)} \right).
		\]
		Thus we have 
		\[
		\left\| \boxtimes^N \Phi - \widetilde{\mathcal{M}}(\Phi)\right\| _\diamond \leq d^n \sqrt{ \frac{2}{d^r} \sum_{\vec{x}: w\left(\vec{x}\right) \notin \mathcal{S}_{\Phi}} \vert{}\Xi_{J_\Phi}(\vec{x})\vert{}^{2N+2}} \leq d^n\sqrt{ \sum_{\vec{x}: w\left(\vec{x}\right) \notin \mathcal{S}_{\Phi}} \vert{}\Xi_{J_\Phi}(\vec{x})\vert{}^{2}\lambda^N },
		\]
		which implies that we obtain a tighter upper bound than the one in Theorem 120 of \cite{Bu3}.
		
		\textbf{(Comparison with the relative entropy-induced Q-CLT)} For the relative entropy induced Q-CLT for quantum states, if we apply the same inequality bounding techniques for quantum channels, we obtain that
		\[
		\mathcal{D}_{Rz}\left(\boxtimes ^{N} \Phi \middle\| \widetilde{\mathcal{M}}\left(\Phi\right)\right) \leq 
		\log \left\lbrace 1+ \frac{\lambda^N}{d^r} \sum_{\vec{x}: w\left(\vec{x}\right) \notin \mathcal{S}_{\Phi}} \vert{}\Xi_{J_\Phi}(\vec{x})\vert{}^{2} \right\rbrace.
		\]
		Thus it holds that
		\[
		\log \left\lbrace 1+ \frac{1}{d^r} \sum_{\vec{x}: w\left(\vec{x}\right) \notin \mathcal{S}_{\Phi}} \vert{}\Xi_{J_\Phi}(\vec{x})\vert{}^{2N+2} \right\rbrace \leq  \log \left\lbrace 1+ \frac{\lambda^N}{d^r} \sum_{\vec{x}: w\left(\vec{x}\right) \notin \mathcal{S}_{\Phi}} \vert{}\Xi_{J_\Phi}(\vec{x})\vert{}^{2} \right\rbrace,
		\]
		where the inequality saturates iff $\vert{}\Xi_{J_\Phi}(\vec{x})\vert{}^2 = \lambda$ for all $\vec{x}$ satisfying $0 < \vert{}\Xi_{J_\Phi}(\vec{x})\vert{} < 1$.
		This implies that the upper bound based on Ruzsa divergence for quantum channels is strictly tighter than the entropic case for quantum states. $\hfill\qedsymbol$
	\end{widetext}

\end{document}